\documentclass[aps,floats,superscriptaddress,twocolumn]{revtex4-2}
\usepackage{graphicx,epsfig}
\usepackage{graphics,dcolumn,bm,epic, eepic,float}
\usepackage{amssymb,amsmath,amsfonts,multirow,rotate,color}
\usepackage{soul,xcolor}

\newcommand{\avg}[1]{{\langle #1 \rangle}}

\begin{document}

\title[Diffusion and synchronization dynamics reveal the multi-scale segregation]{Diffusion and synchronization dynamics reveal the multi-scale patterns of spatial segregation}

\author{Aleix Bassolas}
\thanks{aleix.bassolas@gmail.com}
\affiliation{Departament d'Enginyeria Informatica i Matematiques, Universitat Rovira i Virgili, 43007 Tarragona, Spain}
\author{Sergio G\'omez}
\affiliation{Departament d'Enginyeria Informatica i Matematiques, Universitat Rovira i Virgili, 43007 Tarragona, Spain}
\author{Alex Arenas}
\affiliation{Departament d'Enginyeria Informatica i Matematiques, Universitat Rovira i Virgili, 43007 Tarragona, Spain}
\date{\today}

\begin{abstract}

Urban systems are characterized by po\-pu\-la\-tions with heterogeneous characteristics, and whose spatial distribution is crucial to understand inequalities in life expectancy or education level. Traditional studies on spatial segregation indicators focus often on first-neighbour correlations but fail to capture complex multi-scale patterns.
In this work, we aim at characterizing the spatial distribution heterogeneity of socioeconomic features through diffusion and synchronization dynamics. In particular, we use the time needed to reach the synchronization as a proxy for the spatial heterogeneity of a socioeconomic feature, as for example, the income. Our analysis for 16~income categories in cities from the United States reveals that the spatial distribution of the most deprived and affluent citizens leads to higher diffusion and synchronization times. By measuring the time needed for a neighborhood to reach the global phase we are able to detect those that suffer from a steeper segregation. Overall, the present manuscript exemplifies how diffusion and synchronization dynamics can be used to assess the heterogeneity in the presence of node information.

\tiny
\end{abstract}

\maketitle

\section{Introduction}

The expansion of urbanization and progressive increase of the population in cities has intensified the concern over the many dimensions of segregation ---i.e., school, economic or ethnics--- that have a tangible impact in the health, education and equal opportunities of citizens \cite{kennedy1998income,Elliott1999,collins2000residential,ross2001income,mayer2002economic,acevedo2003residential,wheeler2006urban,owens2018income}. In fact, quantifying the extent of segregation and the identification of economically and socially isolated neighborhoods has been a topic of wide interest that first led to the development of global metrics, and which were later extended to spatial metrics \cite{Cliff1981,Dawkins2004,Brown2006,Dawkins2006,Wong2011,Rey2013}. Most of the initial spatial measures were limited to first neighbour indices, which facilitated the development of multi-scalar indices that provide a more nuanced picture of segregation  \cite{Farber2012,Louf2016,chodrow2017structure,Olteanu2019,sousa2020quantifying,bassolas2021first,bassolas2021diffusion}, yet understanding the role played by each of the scales and their interplay still remains a challenge.

Dynamical processes in general, and in particular diffusion \cite{gomez2013diffusion,sole2013spectral,de2013mathematical,li2013influence,delvenne2015diffusion,de2017diffusion,masuda2017random,cencetti2019diffusive,bertagnolli2021diffusion} and synchronization \cite{arenas2006bsynchronization,arenas2006synchronization,gomez2007paths,gomez2007synchronizability,arenas2008synchronization} dynamics, have been widely studied in complex networks on account of their relation with the spread of diseases and information \cite{gomez2018critical,zhang2016dynamics} and real-world phenomena in social or economic systems \cite{pluchino2005changing,calderon2007trade,erola2012modeling}. Interestingly, they provide insights on the topological scales and structure of networks and reveal the existence of functional meso-scale structures \cite{de2017diffusion,bertagnolli2021diffusion,arenas2006synchronization,gomez2007synchronizability,motter2005network}.

Here we use previous knowledge on diffusion and synchronization dynamics to assess the multi-scale patterns of residential segregation. By moving the focus from the network topology and organization to the node states, we are able to measure how well distributed a population with a certain characteristic is using the time needed to reach the absorbing state. 
Our framework requires thus the implementation of a population dynamic to drive the system towards the homogeneous state, in our case diffusion and synchronization dynamics. None of them constitute here an attempt to model or predict the changes in the spatial distribution of a population characteristic but are highly stylized simplifications of their evolution that allow us to measure the time needed to attain the homogeneous state, which we consider to be the non-segregated scenario. Dynamical approaches are thus introduced here not because they provide a realistic approximation to the evolution of population dynamics but because they offer a significant advantage to measure multi-scale correlations as they do not require to take distance explicitly into account. Moreover, the assumption that cities converge towards uniformity is rather unrealistic without a heavy external driver, and is only a means to construct our measures.

As case studies we provide an analysis on the distribution of citizens of a certain income category in cities from the United States, and the distribution of a set of socioeconomic indicators in the city of Paris throughout an average day (see Supplementary Material Section~2 and Supplementary Figs.~S8-S10). The analysis on the spatial organization of income categories reveals that the most deprived and affluent sectors display higher diffusion and synchronization times linked to a higher heterogeneity, and allow us to split the cities in two groups depending on the difference on the level of segregation. Finally, we evaluate the level of synchronization at the neighborhood level which allow us to spot the more sensitive places in a city.

\section{Results}

\subsection{Diffusion dynamics and income segregation}

Citizens exhibit a huge diversity of characteristics usually captured by socioeconomic indicators such as education level, income or ethnicity, and they are often heterogeneously distributed in space: those individuals with similar characteristics tend to live close between them. To assess the heterogeneity of a population with a characteristic $k\in K$, we consider a graph $G(V,E)$ with adjacency matrix $A=\{a_{ij}\}$ in which the spatial units are represented as a set of nodes $V$ connected by a set of edges $E$. The adjacency matrix $A$ we have considered takes $a_{ij}=1$ when spatial units $i$ and $j$ are adjacent and $a_{ij}=0$ otherwise, which is the traditional connectivity matrix used to capture residential segregation. Still, other types of (weighted) matrices could be considered to assess, for example, the impact of mobility in segregation. The state of a node $x_i^k$ is given by the fraction of citizens living in node $i$ that belong to socioeconomic category (or class) $k$, written as
\begin{equation}
  x_i^k=\frac{n_i^k}{\sum\limits_{k'} n_i^{k'}},
\end{equation}
where $n_i^k$ is the total number of citizens in unit~$i$ that belong to category~$k$. As extreme cases, $x_i^k=0$ when there are no citizens of category~$k$ living in $i$, and $x_i^k=1$ when all the citizens in node~$i$ belong to category~$k$. Of course, the normalization condition
\begin{equation}
  \sum_{k\in K} x_i^k = 1
\end{equation}
is fulfilled for all nodes~$i$.

To measure the multi-scalar patterns of segregation, our assumption is that cities suffering from stronger residential segregation are further from the stationary state where the citizens of category $k$ are homogeneously distributed in space. Although cities are in continuous change and most likely far from equilibrium, similar approaches such as the long-standing Schelling and the Alonso-Muth-Mills models have been able to draw relevant conclusions from the equilibrium state \cite{schelling1971dynamic,fujita1989urban}.

By adopting diffusion dynamics we do not refuse the high complexity of population dynamics influenced by a wide variety of demographic, economic, political, and behavioral factors \cite{zhang2004dynamic,clark2009changing,zhang2011tipping,deluca2013segregating} but avoid introducing further parameters and factors that could hinder our aim of characterizing the segregation of a particular population category. Bear in mind that our final goal is by no means to assess real-world migration processes but to construct a multi-scalar measure of segregation that does not explicitly include the distance and the use of more complex and realistic approaches that would complicate the interpretation of the results. Diffusion constitutes one of the most basic approximations to how information, or any other characteristic, is transmitted through a system. Although far from the real behavior, it provides one of the simplest scenarios where the flow of population follows a gradient.

In fact, we focus on one of the best-case scenarios where the values of $x_i^k$ converge towards equilibrium following a gradient, which could be interpreted as the change of residence of citizens of category $k$ to regions where they are less abundant.

\begin{table}[t!]
\centering
\begin{tabular}{lc}
  \hline
  \hline
  Class & Income (\$) \\
  \hline
   1 & Less than 10,000 \\
   2 &  10,000 --  14,999 \\
   3 &  15,000 --  19,999 \\
   4 &  20,000 --  24,999 \\
   5 &  25,000 --  29,999 \\
   6 &  30,000 --  34,999 \\
   7 &  35,000 --  39,999 \\
   8 &  40,000 --  44,999 \\
   9 &  45,000 --  49,999 \\ 
  10 &  50,000 --  59,999 \\
  11 &  60,000 --  74,999 \\
  12 &  75,000 --  99,999 \\
  13 & 100,000 -- 124,999 \\
  14 & 125,000 -- 149,999 \\ 
  15 & 150,000 -- 199,999 \\
  16 & 200,000 or more    \\
  \hline 
  \end{tabular}
  \caption{Income range (in US dollars) corresponding to each category (or class).} \label{Table1}
\end{table}

 \begin{figure*}[t!]
    \begin{center}
    \includegraphics[width=0.95\textwidth]{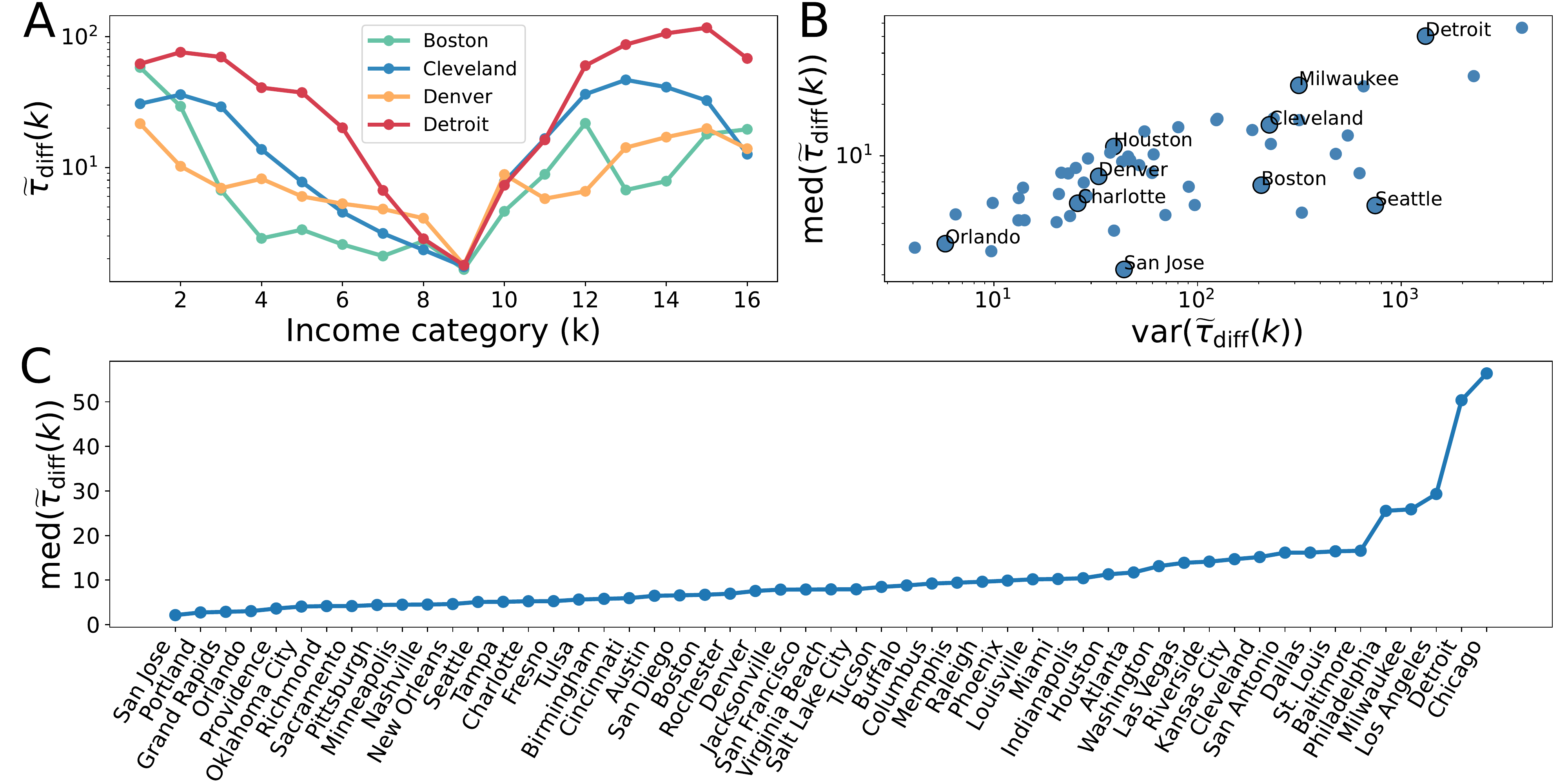}
    \end{center}
    \caption{\textbf{Diffusion dynamics as a measure for income segregation.} (A) Synchronization time for each of the $16$ income categories in Boston, Cleveland, Denver and Detroit. (B) Median value of $\widetilde{\tau}_{\rm diff}(k)$ across income categories as a function of its variance. (C) Ranking for the median value of  $\widetilde{\tau}_{\rm diff}(k)$ for the studied set of US cities. }
    \label{fig1}
\end{figure*}

\begin{figure}[t!]
    \begin{center}
    \includegraphics[width=0.999\columnwidth]{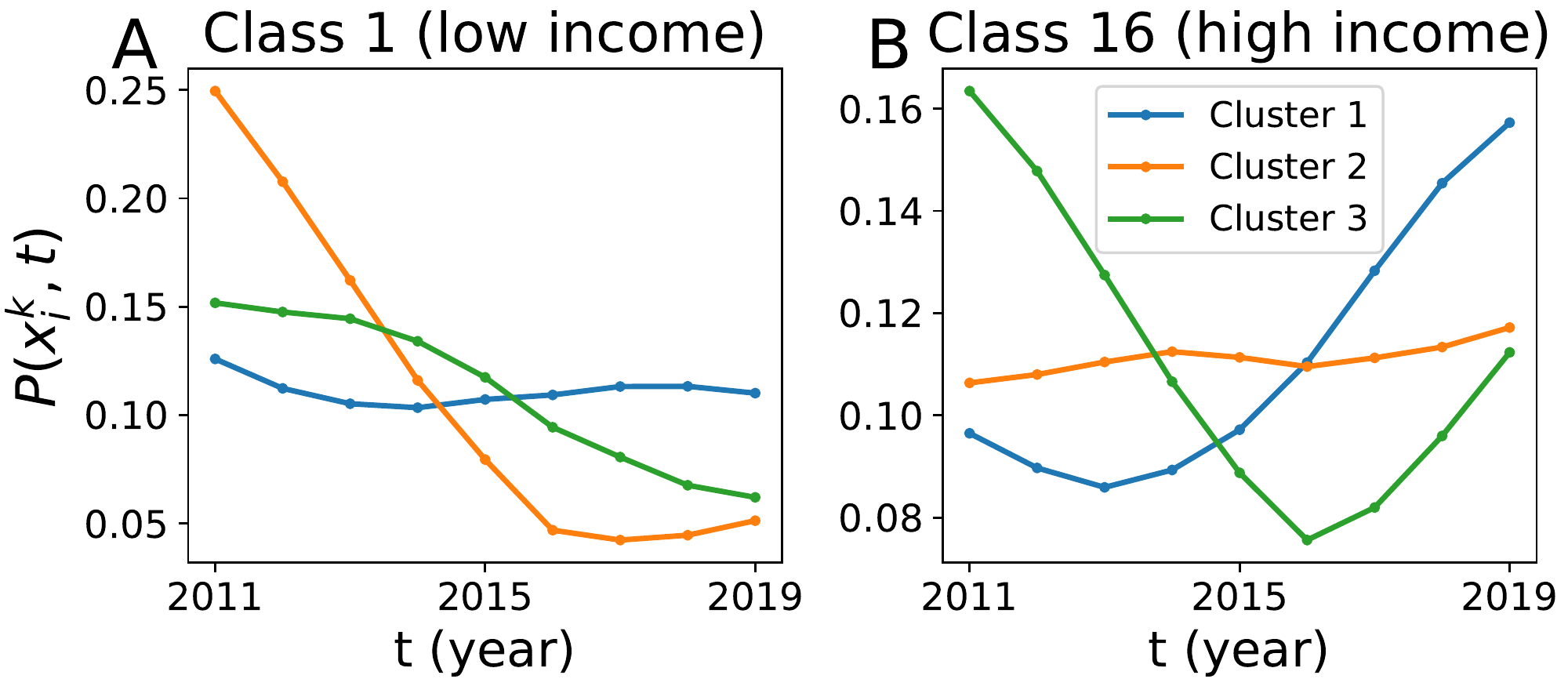}
    \end{center}
    \caption{\textbf{Average temporal evolution of the abundance of households within the lowest and highest income.} Temporal evolution of centroids after performing a k-means clustering on the normalized abundance of households with category~$k$, $P(x_i^k,t)$, as a function of time~$t$ for the lower (A) and higher (B) income categories.}
    \label{fig2}
\end{figure}

 \begin{figure*}[t!]
    \begin{center}
    \includegraphics[width=0.90\textwidth]{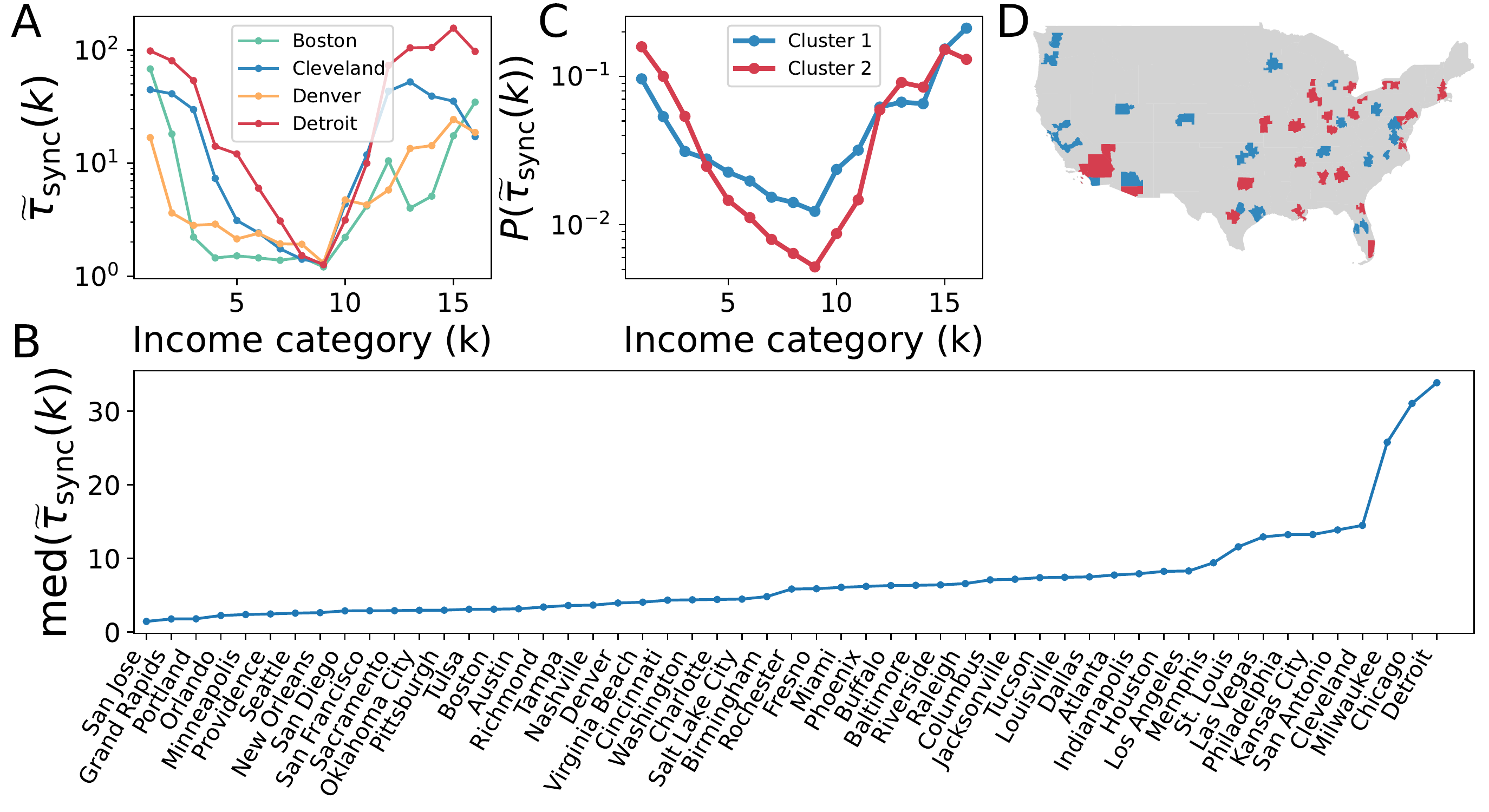}
    \end{center}
    \caption{\textbf{Synchronization time as a measure for income segregation.} (A) Synchronization time for each of the $16$ income categories in Boston, Cleveland, Denver and Detroit. (B) Ranking for the median value of $\widetilde{\tau}_{\rm sync}(k)$ for the studied set of US cities. (C) Average value of $P(\widetilde{\tau}_{\rm sync}(k))$ as a function of each income category~$i$ for the two main clusters detected. (D) Location and cluster assignment for each of the analyzed cities.}
    \label{fig3}
\end{figure*}

We focus on the economic segregation in the metropolitan areas of the United States with more than $1$~million inhabitants and analyze a dataset containing the number of households within an income interval~$k$ residing in each census tract (see Table~\ref{Table1}). 

Once we have the set of initial node states $x_i^k$, their evolution through time is determined by the diffusion dynamics
\begin{equation}
  \frac{dx_i^k}{dt}=\frac{1}{s_i}\sum\limits_{j=1}^N a_{ij}(x_j^k-x_i^k),
\end{equation}
where
\begin{equation}
  s_i = \sum\limits_{j=1}^N a_{ij}
\end{equation}
is the degree of node $i$. For simplification purposes, we have opted to use a normalized diffusion dynamic, with diffusion strength equal to~1. Note that we have independent diffusion processes for each category~$k$.

The diffusion dynamic lasts until the stationary state, $x_i^k=\avg{x^k}$, $\forall i$ is reached, and we denote the spanned time as $\tau_{\rm diff}(k)$. Since the time to reach the stationary state can be infinitely large, we have considered that it is reached when the variance of $x_i^k$, in time, becomes lower than $0.0001$. We hypothesize that lower values of $\tau_{\rm diff}(k)$ are related to a more homogeneous distribution of the population within a category~$k$, and the other way around when it is higher. In the extreme case in which all units have the same initial value of $x_i^k$, the diffusion time $\tau_{\rm diff}(k)$ would attain its minimum value. As we aim to compare cities with different characteristics, we control for confounding factors such as the particular distribution of $x^k$ or the topology of the graph by running the same diffusion dynamics on the same graph but where the values of $x^k$ have been reshuffled, thus defining the average null-model diffusion time $\tau^{\rm null}_{\rm diff}(k)$ calculated over $500$ reshuffling realizations. The relative diffusion time we will use throughout this manuscript can then be written as
\begin{equation}
  \widetilde{\tau}_{\rm diff}(k) = \frac{\tau_{\rm diff}(k)}{\tau^{\rm null}_{\rm diff}(k)}.
\end{equation}
A relative diffusion time equal to one means that
it is compatible with the null model, i.e., there are no remarkable spatial dependencies, while a greater value suggests that spatial heterogeneities delay the arrival to the stationary state.

We analyze the normalized diffusion times $\widetilde{\tau}_{\rm diff}(k)$ by running simulations for all US cities above $1$ million of inhabitants and each of the $16$ income categories~$k$ as a proxy for how heterogeneously distributed is the population; we have excluded New York City, whose adjacency network does not provide an accurate picture of residential segregation due to the particular geography of Manhattan. In Fig.~\ref{fig1}(A) we display $\widetilde{\tau}_{\rm diff}(k)$ in Boston, Cleveland, Detroit and Denver observing a common qualitative behavior: smaller values for middle-income categories, and higher ones for the categories in the extremes of the income distribution. Our results suggest that the wealthier and most deprived citizens suffer from stronger segregation and display a more clustered spatial distribution. More interestingly, category~$9$ seems to be the more homogeneously distributed across space, in agreement with the results observed in \cite{Bassolas2020b} and with the mean and standard deviation of $x^k$ as well as the Moran's I (see Supplementary Material Section~1 and Supplementary Fig.~S1). Still, there are strong quantitative differences, with Cleveland and Detroit displaying higher values for most of the categories, in contrast to Boston and Denver.

Since $\widetilde{\tau}_{\rm diff}(k)$ takes a set of $16$~values for each city, we calculate their median and variance values over all categories to ease the comparison between the set of cities studied. While the median value provides information on the segregation across all economic categories, its variance reports the variability among them. Figure~\ref{fig1}(B) shows this median value of $\widetilde{\tau}_{\rm diff}(k)$, $\mbox{med}(\widetilde{\tau}_{\rm diff}(k))$ as a function of its variance, $\mbox{var}(\widetilde{\tau}_{\rm diff}(k))$. The prior cities appear ordered as Detroit, Cleveland, Boston and Denver, although the variance is very similar for Cleveland and Boston, likely due to the high values observed for low-income categories in Boston. Finally, we provide in Fig.~\ref{fig1}(C) the ranking of the selected US cities according to $\mbox{med}(\widetilde{\tau}_{\rm diff}(k))$, as a measure of the overall segregation in cities. On top of it, we find cities such as Milwaukee or Detroit, which have been reported to suffer from economic and ethnic segregation \cite{adelman2004neighborhood,thomas2015race,florida2015segregated}. 

By applying diffusion dynamics we implicitly assume that $x^k$ evolves homogeneously towards consensus, which more than a realistic scenario, it is a means to calculate the time needed to reach consensus and obtain a measure of segregation. To further inspect the actual change of $x_i^k$ between 2011 and 2019 in each of the spatial units~$i$, we first construct the normalized time-series for each spatial unit across those years as
\begin{equation}
  P(x_i^k,t)=\frac{x_i^k(t)}{\sum\limits_{t'} x_i^k(t')},
\end{equation}
and then cluster, for each category~$k$, the temporal profiles of all the nodes. For the clustering, we have made use of the k-means algorithm \cite{hartigan1979algorithm,likas2003global}, grouping together those units with a similar temporal evolution, and setting the number of clusters to~3. The resulting time-series of the corresponding centroids for the highest and lowest income categories are depicted in Fig.~\ref{fig2}, where a non-monotonic behavior is observed in most of the cases, with oscillatory behaviors through time of varying amplitude. 

\begin{figure*}[t!]
    \begin{center}
    \includegraphics[width=0.85\textwidth]{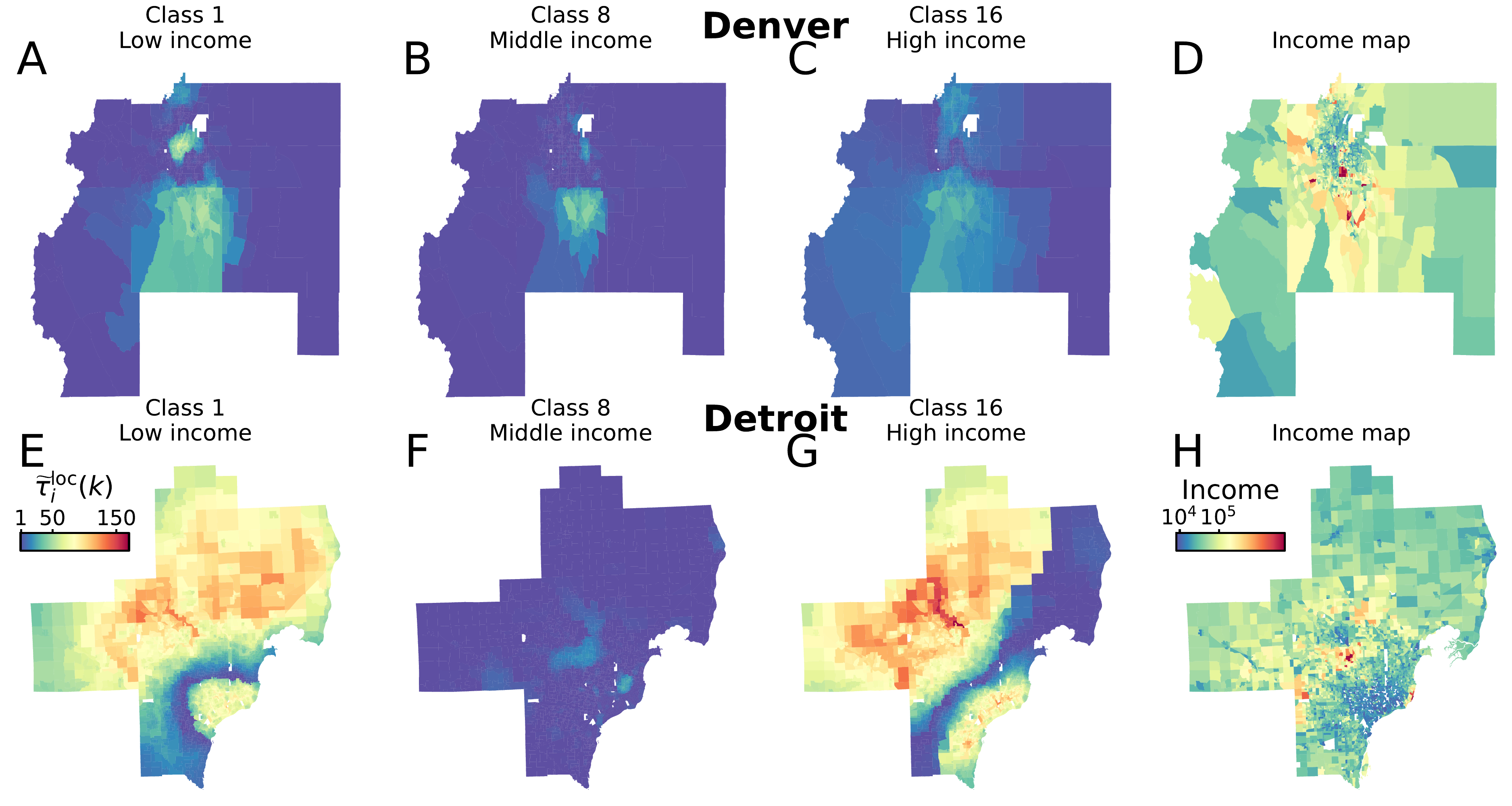}
    \end{center}
    \caption{\textbf{Local synchronization time as a measure for income segregation.}  Normalized synchronization time for each census tract in Denver (A--C) and Detroit (E--G) for three different income categories: (A, E) class~1 (low income); (B, F) class~8 (middle income); (C, G) class~16 (high income). We provide as a reference the median income of each census tract in (D) Denver and (H) Detroit.}
    \label{fig4}
\end{figure*}

\subsection{Synchronization dynamics and income segregation}

According to the oscillations in the temporal evolution of $x^k_i$ (Fig.~\ref{fig2}), diffusion dynamics appear to be a rather simplistic approach to assess the time needed to converge. Even thought we do not aim to mimic the real evolution of $x^k_i$, we seek for a dynamic that at least can resemble its real behaviour in a qualitative way. Thus, despite still constituting a stylized approximation, a dynamical process with an oscillatory behavior, like a system of coupled Kuramoto oscillators, appears to be a better way to assess the spatial heterogeneity of socioeconomic indicators across cities. To analyze segregation in terms of synchronization dynamics, we treat each of the spatial units~$i$ as an individual Kuramoto oscillator, with an initial phase $\theta_i^k(0)$ that is set by distributing the fraction of population in node $i$ that belongs to a category~$k$ within the range $[0,\pi]$ as
\begin{equation}
  \theta_i^k(0)=x_i^k\pi.
\end{equation}
The interaction between spatial units is given by the Kuramoto model
\begin{equation}
  \dot{\theta}_i^k(t)=\omega_i^k+\frac{1}{s_i}\sum\limits_{j=1}^N a_{ij} \sin\left(\frac{\theta^k_j(t)-\theta_i^k(t)}{2}\right),
\end{equation}
where we have modified the traditional interaction term between oscillators by dividing the angle difference by two, allowing for the interaction between regions displaying extreme values of $x_i^k$. Additionally, to facilitate the global synchronization of the system, we set all the individual natural frequencies of the oscillators to the same value, i.e., $\omega_i=1,\ \forall i$. In order to account separately the segregation of each category $k$, our approach assumes that there is no interaction between categories and, thus, $x_i^k$ synchronize independently of $k$.

\begin{figure*}[t!]
    \begin{center}
    \includegraphics[width=0.80\textwidth]{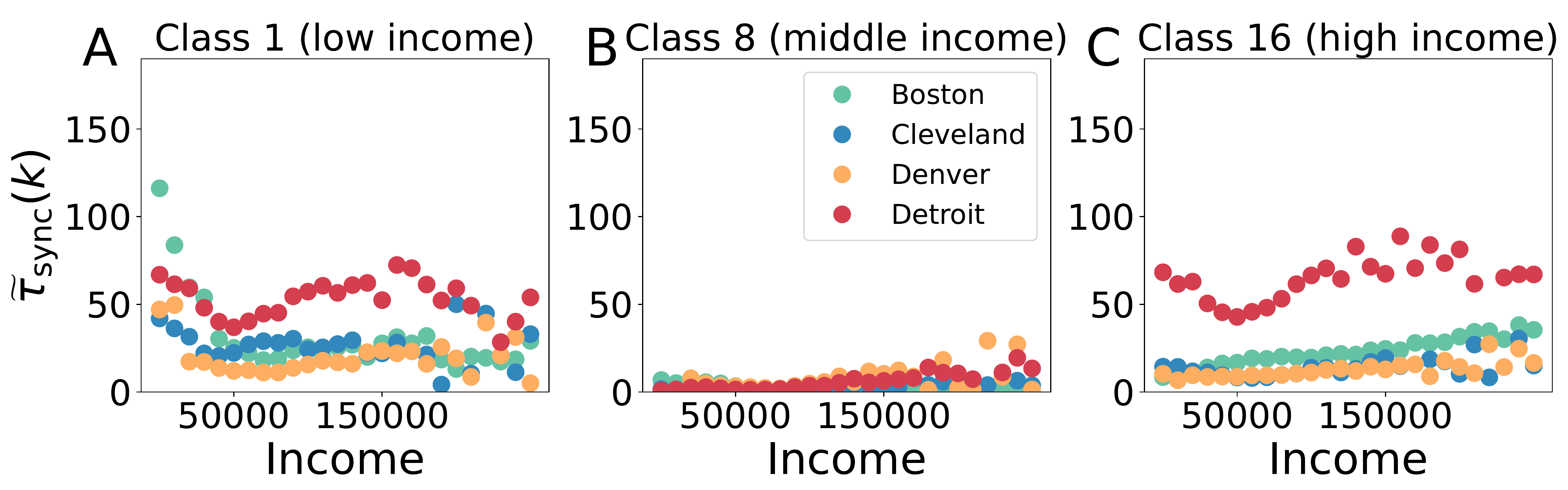}
    \end{center}
    \caption{\textbf{Synchronization time and median income.} Normalized synchronization time as a function of the median income averaged over bins of \$5,000 for class~1 (A), class~8 (B) and class~16 (C).}
    \label{fig5}
\end{figure*}

We use the standard order parameter $|z^k|$ to assess the global level of synchronization for a category $k$ in a city, where
\begin{equation}
  z^k=\frac{1}{N}\sum\limits_{j=1}^N  e^{i\theta_j^k},
\end{equation}
and $N$ is the total number of spatial units or Kuramoto oscillators \cite{arenas2008synchronization}. We consider that a city has reached the synchronized state when $|z^k|>0.999$. As in the case of diffusion, we assess how the distribution of initial phases determines the synchronization of the system, a city in our case, by measuring the time $\tau_{\rm sync}(k)$ required to reach the synchronized state. The more heterogeneously distributed the initial phases are, the higher the time the system requires to synchronize. To distinguish between the effect produced by the spatial distribution $x_i^k$ from its overall distribution as well as the topology of the graph, we also measure the average time the system needs to synchronize when the same phases are redistributed at random, $\tau^{\rm null}_{\rm sync}(k)$. The normalized synchronization time of the system is then given by the ratio 
\begin{equation}
  \widetilde{\tau}_{\rm sync}(k)=\frac{\tau_{\rm sync}(k)}{\tau^{\rm null}_{\rm sync}(k)}.
\end{equation}
Like for diffusion, a synchronization time close to one means that the spatial distribution of phases is compatible with the null model, and a larger value indicates that spatial heterogeneities delay the appearance of a synchronized state.

In Fig.~\ref{fig3}(A) we inspect the normalized synchronization time in Boston, Cleveland, Detroit and Denver when spatial units interact through Kuramoto-like dynamics. All four of them share similar features, with central classes displaying smaller synchronization times compared to the most disadvantaged and wealthier ones. An expected result since those individuals in the extremes of the income distribution tend to be more isolated and clustered together compared to middle-income citizens. Despite sharing qualitative features, the cities shown display sharp quantitative differences. Almost all categories appear to be significantly more isolated in Detroit and Cleveland compared to Denver and Boston, where $\widetilde{\tau}_{\rm sync}(k)$ looks much flat. Overall, the synchronization results are compatible with the diffusion ones, likely because both dynamical processes share common features. We have further checked that the mean $\avg{x^k}$ does not determine directly the normalized synchronization times in Supplementary Fig.~S1.

Likewise with diffusion, we calculate the median and variance of $\widetilde{\tau}_{\rm sync}(k)$ over all categories to be able to compare between analyzed cities (see Supplementary Fig.~S2 for the individual rankings of $\widetilde{\tau}_{\rm sync}(k)$ for the categories $k=1$ and $k=16$). The ranking is shown in Fig.~\ref{fig3}(B) and has cities such as Detroit, Cleveland, Milwaukee or Memphis close to the top, which are well-known for being among the most economically segregated cities in the United States. The location in the ranking of the cities in Fig.~\ref{fig3}(A) is consistent with our observations, with Boston and Denver on the bottom of the ranking and Detroit and Cleveland on the top of it.

Our index is given by the median value of the normalized synchronization times, yet depending on the dimension of segregation we aim to capture, we can also construct an index based on a population-weighted average. Whereas the median gives equal weight to each economic category focusing on the segregation suffered by residents of category $k$, the weighted average provides an overall picture of segregation taking the population of each category into account. We show the ranking obtained for the weighted average index and its relation with $\mbox{med}(\widetilde{\tau}_{\rm sync}(k))$ in Supplementary Figs.~S6 and~S7. Additionally, we show in Supplementary Figs.~S3 and~S4 how $\widetilde{\tau}_{\rm sync}(k)$ significantly correlates with the traditional Moran's I \cite{moran1948interpretation} as well as a multi-scale quantity based on class mean first passage times developed in \cite{bassolas2021first,bassolas2021diffusion}, reinforcing the idea that synchronization (and diffusion) dynamics indeed capture the patterns of residential segregation. Despite the dynamics we have used are stylized versions of the real behavior of the quantity $x^k_i$ and do not capture the full complexity of its temporal evolution, it is able to capture segregation with values comparable to other segregation indicators.

Although $\widetilde{\tau}_{\rm sync}(k)$ is larger for extreme categories in most of the cities, some of them like Denver display smaller variations than others such as Detroit and, therefore, it might be of interest to group cities according to the change in synchronization times. By running a k-means algorithm on the normalized value of $P(\widetilde{\tau}_{\rm sync}(k))$ so that $\sum_{k} P(\widetilde{\tau}_{\rm sync}(k))=1$, we can split the cities of study between those with higher and smaller differences in $\widetilde{\tau}_{\rm sync}(k)$, see Fig.~\ref{fig3}(C). In Fig.~\ref{fig3}(D), we display the cluster assigned to each metropolitan area, where no strong spatial pattern is observed. Still, the cities in the Midwest, which are known for being economically segregated, fall into the red cluster, together with other cities such as Baltimore or Los Angeles. If, instead, we focus on the blue cluster, we have cities such as Sacramento or Washington D.C. Among the cities discussed in Fig.~\ref{fig3}(A), Denver falls into the group with more homogeneous segregation (in blue) and the rest into the one with more unequal segregation patterns (in red). 
 
Beyond the global quantification of segregation, we can also evaluate the local level of segregation of a concrete census tract $i$  at a given time step $t$ by computing
\begin{equation}
  \rho^k_i(t)=\cos(\theta^k_i(t)-\Phi^k(t)),
\end{equation}
where $\theta^k_i(t)$ is the phase of unit $i$ at time $t$ and $\Phi^k(t)$ is the average phase of all the oscillators in a city in a given time $t$ \cite{arenas2006synchronization}. When $\rho^k_i(t)>0.999$ we consider that oscillator $i$ has synchronized, from which we can obtain $\tau^{\rm loc}_i(k)$. However, given that $\rho_i(\tau^{\rm loc}_i(k))$ can oscillate through time, we only consider that a unit $i$ has reached the global synchronized state at a time $\tau^{\rm loc}_i(k)$ when $\rho_i(t>\tau^{\rm loc}_i(k))$ does not go below $0.999$ anymore, otherwise our methodology could fail to capture long-range correlations. In order to provide a metric for each spatial unit, simulations last until each of the spatial units have fullfiled the synchronization criteria. Normalizing $\tau^{\rm loc}_i(k)$ by its null model counterpart, it yields $\widetilde{\tau}^{\rm loc}_i(k)$, a measure of the local synchronization time.

Figure~\ref{fig4} displays the normalized synchronization times for each of the census tracts in Denver and Detroit, focusing on three very distinct income categories: low income, Fig.~\ref{fig4}(A,E); middle income, Fig.~\ref{fig4}(B,F); and high income, Fig.~\ref{fig4}(C,G). To ease the comparison between income categories, the range of values is common for all the maps, evincing the strong differences between Detroit and Denver, especially for the low and high-income categories. The shape of the segregation in Detroit can be outlined by the lower-income downtown and the richer suburbs, being the most segregated parts, and a less-segregated region in-between. In the case of Denver, we only slightly see high values for the low-income category in the North of the city and the high-income category in the South.

As we detail in Supplementary Fig.~S5, the spatial patterns of segregation product of the synchronization dynamics are significantly different to those obtained from first-neighbor quantities such as the Moran's I. Instead of focusing on those regions whose proportion of citizens is high (or low) compared to its neighbors, our methodology highlights those with a ratio of population within a category~$k$ distinct than the average, either because it is high or low, and spatially isolated from those regions with average values. In other words, a region with a high proportion of residents of category $k$ might not show a large local spatial correlation if their neighbors have similar values but could, instead, produce high values of $\widetilde{\tau}^{\rm loc}_i(k)$ if it is isolated from those regions displaying a proportion of citizens closer to the city average. As the majority of spatial measures, our approach can also suffer the so-called modifiable areal unit problem \cite{fotheringham1991modifiable} in a similar fashion. However, given that our methodology captures mid and long-range correlations instead of local differences, it might be less affected by such small local changes.


Finally, we inspect if the synchronization time of a region displays any type of connection with its actual income. To do so, we plot in Fig.~\ref{fig5} the normalized local synchronization time as a function of the median income averaged over all the census tracts within bins of \$5,000 in four US cities. Again we see that segregation is much stronger in Detroit followed by Cleveland and Boston. High-income regions are more segregated in Boston compared to Cleveland. In general terms, the census tracts with a median income between \$50,000 and \$80,000 seem to be the less segregated ones as they synchronize faster for both low and high-income categories. These results are in agreement with the cluster assignment of the previous cities, with Detroit, Cleveland and Boston in the red cluster where low and high-income categories need more time to synchronize, and Denver in the blue cluster where only the high-income categories need more time to synchronize.

\section{Discussion}

Traditional spatial segregation indicators that focus on local scale of segregation fail in most cases to capture the presence of long-range correlations, thus highlighting the need of multi-scale indices \cite{Farber2012,Louf2016,chodrow2017structure,Olteanu2019,bassolas2021first,
sousa2020quantifying,bassolas2021diffusion}. Our framework does not consider any specific scale, but uses a dynamical approach that captures the patterns of segregation across the multiple scales. We have revealed how categories in the extreme of the income distribution are more heterogeneously distributed in space compared to middle classes, displaying larger diffusion and synchronization times. This approach has also allowed us to group together those cities that display common features of segregation. In this context, it is important to note that our work does not attempt to model the evolution of income segregation nor can be used as a forecasting tool, but takes modeling assumptions to assess the level of segregation that a distribution of population exhibits. 

Despite the main manuscript focuses on the economic segregation, our methodology can be used to assess the heterogeneity in the spatial distribution of any characteristic. Moreover, it can go beyond the spatial component of segregation by including in the analysis other types of graphs, e.g., the daily mobility network of citizens. In this way, we could assess how citizens of diverse socioeconomic environments interact through mobility \cite{xu2019quantifying,toth2021inequality,bokanyi2021universal,moro2021mobility}. 

Summarizing, we show how diffusion and synchronization dynamics can be used in some systems to assess the heterogeneity in the distribution of node features. While the present work focuses on the initial phases of oscillators and their synchronization time, node metadata could also be understood as an internal frequency and provide further insights on feature correlation across topological scales.

\section*{Conflict of Interest Statement}

The authors declare that the research was conducted in the absence of any commercial or financial relationships that could be construed as a potential conflict of interest.

\section*{Author Contributions}
A.B.\ performed the research. A.B.\, S.G.\ and A.A.\ designed the research and wrote the manuscript.

\section*{Acknowledgments}

A.B.\ acknowledges financial support from the Ministerio de Ciencia e Innovaci\'on under the Juan de la Cierva program (FJC2019-038958-I). We acknowledge support by Ministerio de Econom\'{\i}a y Competitividad (PGC2018-094754-BC21, FIS2017-90782-REDT and RED2018-102518-T), Generalitat de Catalunya (2017SGR-896 and 2020PANDE00098), and Universitat Rovira i Virgili (2019PFR-URV-B2-41). A.A.\ acknowledges also ICREA Academia and the James S.\ McDonnell Foundation (220020325).

\section*{Data Availability Statement}
The income data analyzed in the present text can be found at \cite{income}.

\clearpage

\renewcommand\theequation{{S\arabic{equation}}}
\renewcommand\thetable{{Supplementary S\Roman{table}}}
\renewcommand{\figurename}{Supplementary Figure}
\renewcommand\thefigure{{S\arabic{figure}}}
\renewcommand\thesection{{Section S\arabic{section}}}

\setcounter{section}{0}
\setcounter{table}{0}
\setcounter{figure}{0}
\setcounter{equation}{0}

\onecolumngrid


\section{Supplementary results for diffusion and synchronization dynamics and economic segregation}

\begin{figure}[ht!]
    \begin{center}
    \includegraphics[width=0.8\textwidth]{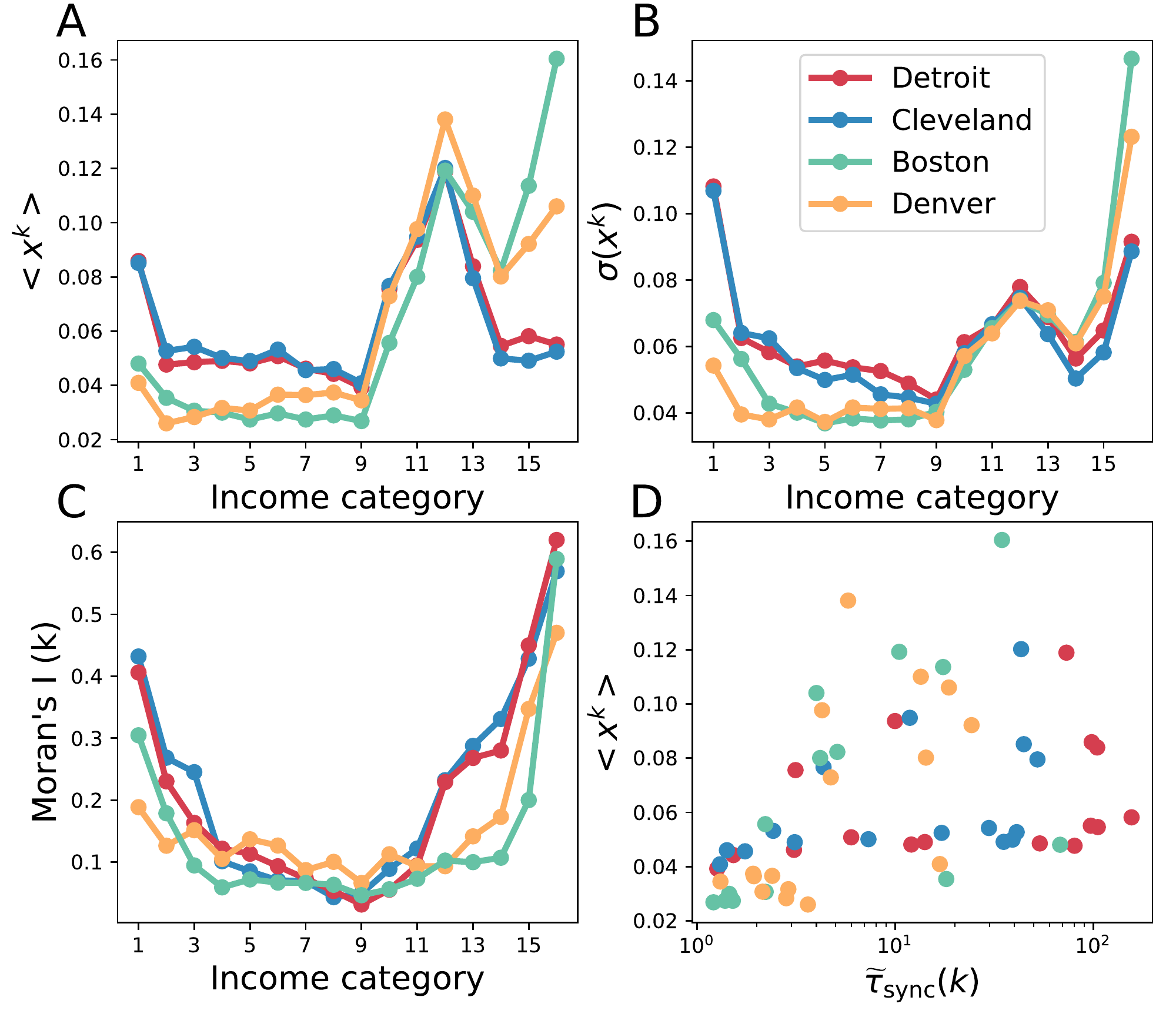}
    \end{center}
      \caption{(A) Mean of $x^k$ for each income category in Boston, Cleveland, Detroit and Denver. (B)~Standard deviation $\sigma(x^k)$ for each income category in Boston, Cleveland, Detroit and Denver. (C)~Moran's I for each income category in Boston, Cleveland, Detroit and Denver. (D)~Scatter plot of the mean of $x^k$ as a function of $\widetilde{\tau}_{\rm sync}(k)$.}
     \label{figS1}
\end{figure}

We provide here supplementary results related to the study of income segregation in US cities. Figure~\ref{figS1} reports (A) the mean and (B) standard deviation of $x_i^{k}$ in Boston, Cleveland, Detroit and Denver. Both of them reach minimum values between 8-10. 

The fact that classes 8-10 appear to be the less segregated is also supported by the Moran's I as  Fig.~\ref{figS1}(C) shows. To further assess that the mean $<x^{k}>$ does not strongly determine the values of $\widetilde{\tau}_{\rm sync}(k)$, we plot both quantities in Fig.~\ref{figS1}(D), where no strong pattern is observed. Categories with low  $<x^{k}>$ display high variability in $\widetilde{\tau}_{\rm sync}(k)$ and vice-versa.

\begin{figure}[ht!]
    \begin{center}
    \begin{tabular}{l}
    A\hfill Class 1 \hfill\mbox{} \\
    \includegraphics[width=0.95\textwidth]{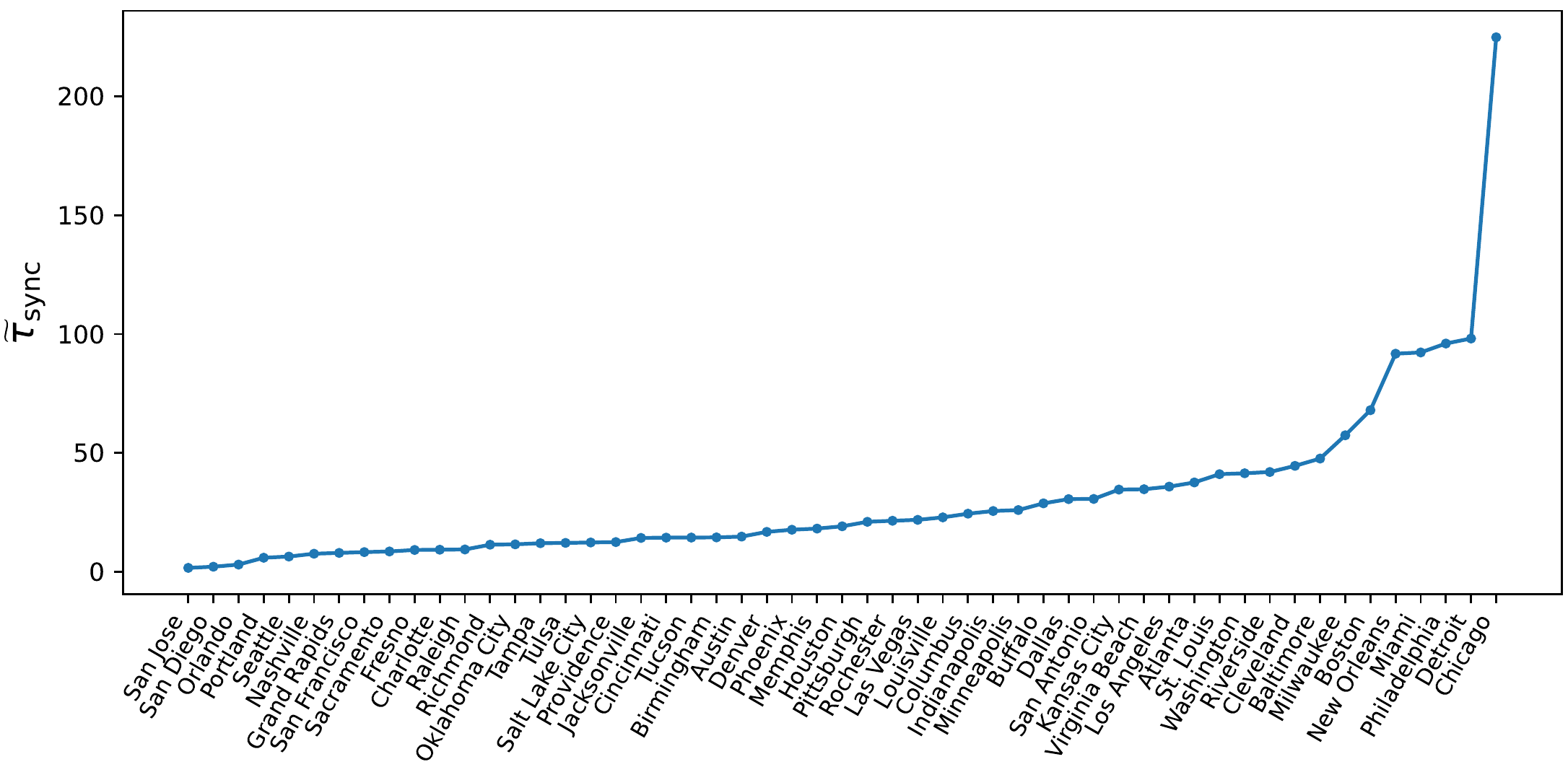} \\
    B\hfill Class 16 \hfill\mbox{} \\
    \includegraphics[width=0.95\textwidth]{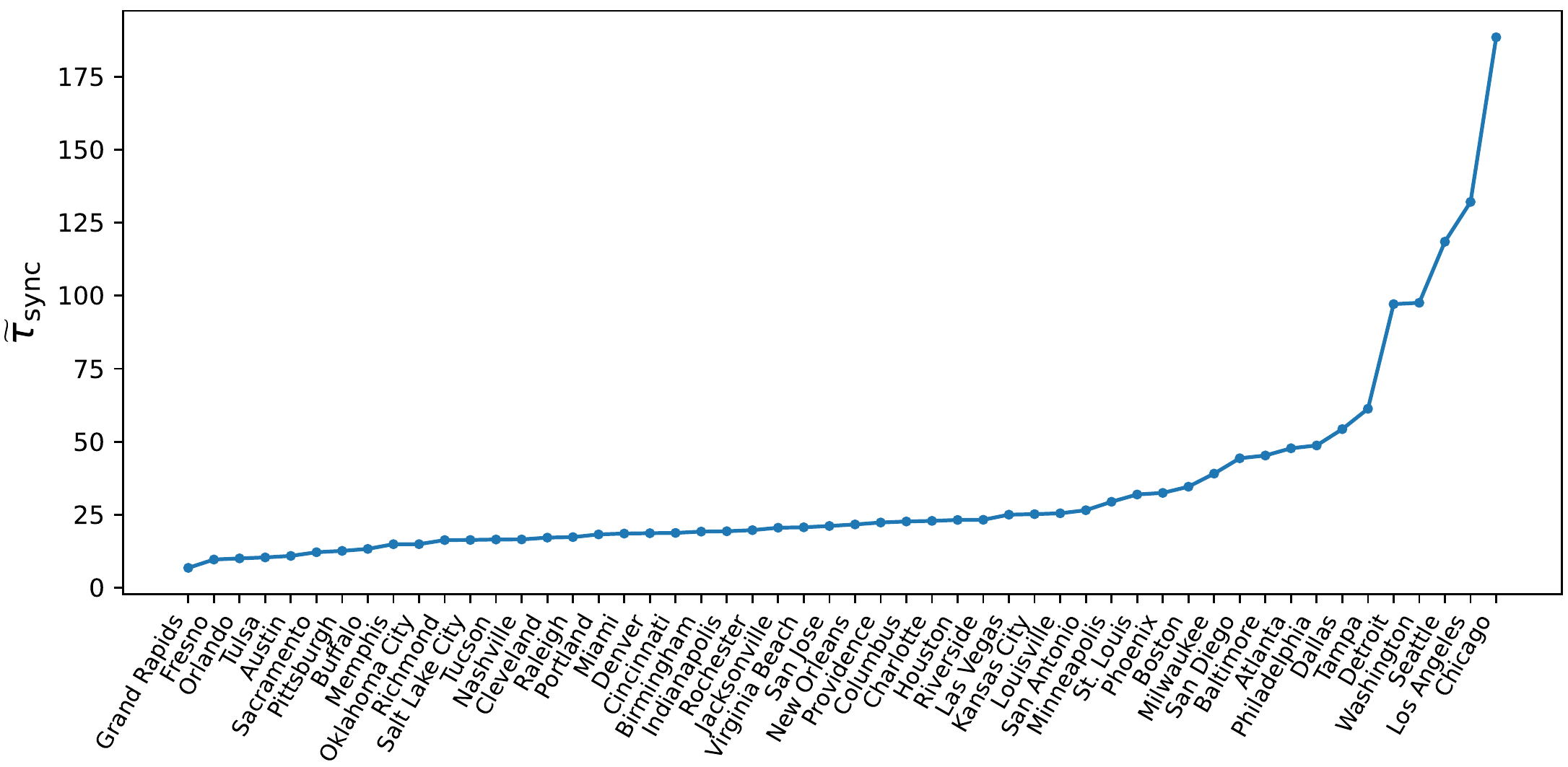}
    \end{tabular}
    \end{center}
    \caption{Ranking of the selected US cities according to the value of $\widetilde{\tau}_{\rm sync}(k)$, for income class~1 (A) and income class~16 (B).}
     \label{figS2}
\end{figure}

In Fig.~\ref{figS2} we provide the ranking of the selected US cities according to the value of $\widetilde{\tau}_{\rm sync}(k)$ for the lowest and highest income categories~1 and 16, respectively. As can be seen, there are significant variations in the ranking depending on which economic category is shown; for example, Cleveland is close to the top for category 1 but far apart for 16, and the other way around for Seattle.

\clearpage

\section{Comparison with other segregation measures}

In this section we assess how the normalized synchronization time $\widetilde{\tau}^k_{\rm sync}$ relates to other segregation measures. In particular we focus on the widely used Moran's I \cite{moran1948interpretation}, which focuses on local correlations, and one obtained from class mean first passage times (CMFPT) developed in \cite{bassolas2021first,bassolas2021diffusion}, which captures long-range spatial correlations. 

For each city and category $k$ the Moran's I can be written as
\begin{equation}
  I^k = 
  \frac{\displaystyle\frac{1}{W}\sum_{i=1}^n \sum_{j=1}^n w_{ij}(x^k_i - \bar{x}^k)(x^k_j -
  \bar{x}^k)}{\displaystyle\frac{1}{n}\sum_{i=1}^n (x^k_i - \bar{x}^k)^2},\label{eq:morani}
\end{equation}
where $x^k_i$ is the fraction of population in $i$ that belongs to category $k$, $\bar{x}^k$ is its mean across all spatial units, the weights $w_{ij}$ correspond in our case to the spatial adjacency matrix $a_{ij}$, and $W=\sum_{i=1}^n\sum_{j=1}^n w_{ij}$ is the total weight.
 
As an index to assess the long-range correlations in the spatial distribution of the income categories, we will use the class mean first passage times between classes. In this methodology \cite{bassolas2021first,bassolas2021diffusion}, random walkers start from each of the spatial units in a system and move through the spatial adjacency graph until they have visited the $16$~classes at least once. For this, each location is assigned to a class with probability proportional to its corresponding fraction of population. By averaging the number of steps that a walker needs to reach class $j$ across all the units that belong to category $i$ and for multiple realizations, we can obtain the class mean first passage times $\tau_{ij}$, which encapsulate the average number of steps needed to reach a unit of category $j$ when a walker departs from a unit of category $i$. After normalizing by a null-model in which colors are uniformly reshuffled at random to compensate for uneven class abundances, we finally obtain the normalized class mean first passage times $\widetilde{\tau}_{ij}$. The quantity $\widetilde{\tau}_{ij}$ provides thus information on how much time you need to reach category $j$ when a walker departs from a unit of category $i$ as compared to the null-model, values below $1$ mean that two categories are closer than in the null-model and vice-versa for values above $1$. To summarize the segregation of category $k$ in a city we will use the CMFPT index, i.e., the $\mbox{med}(\widetilde{\tau})_k$ given by the median value of $\tau_{jk}$ $\forall j$. 

\begin{figure}[bh!]
    \begin{center}
    \includegraphics[width=0.45\textwidth]{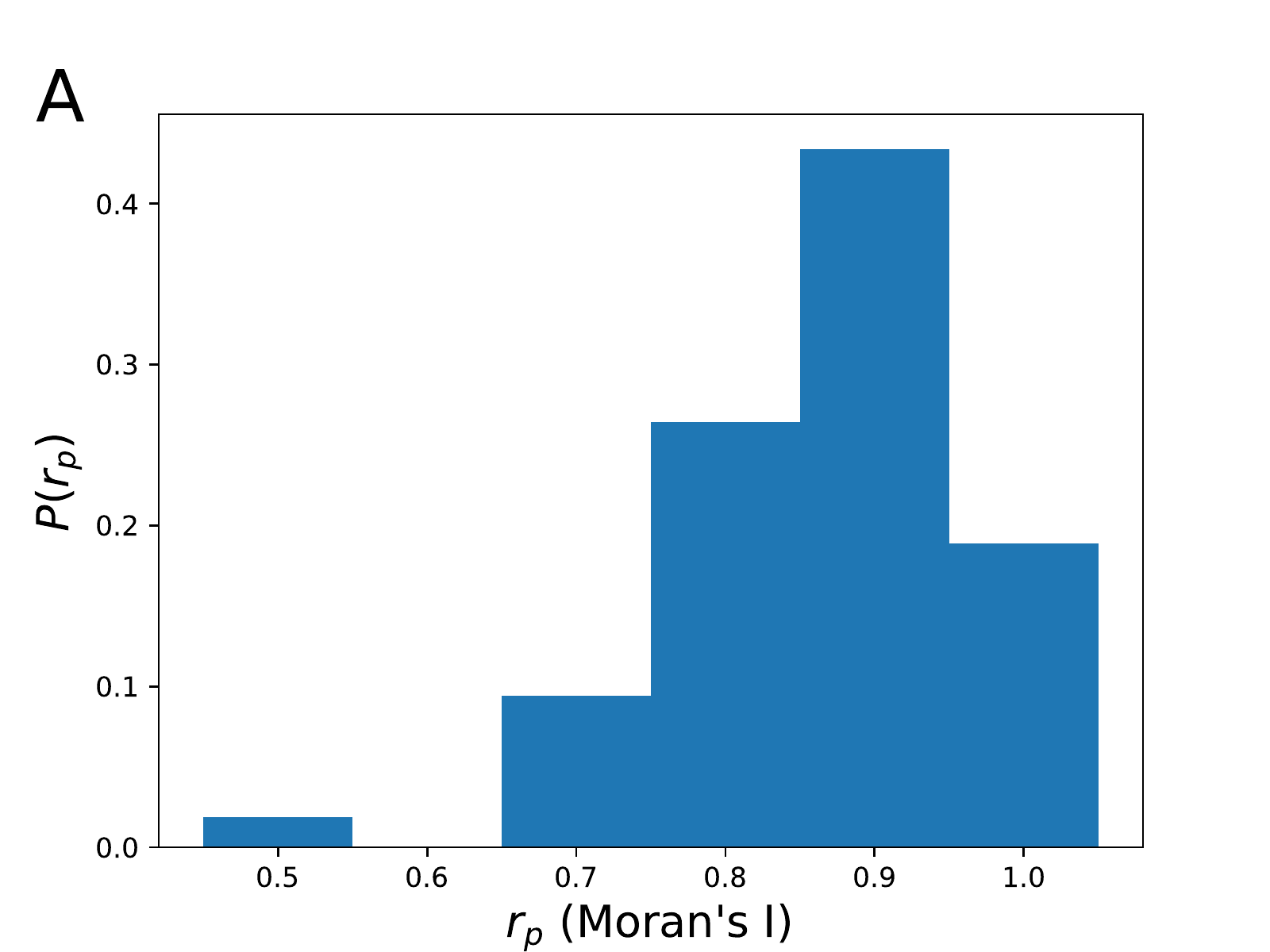}
    \includegraphics[width=0.45\textwidth]{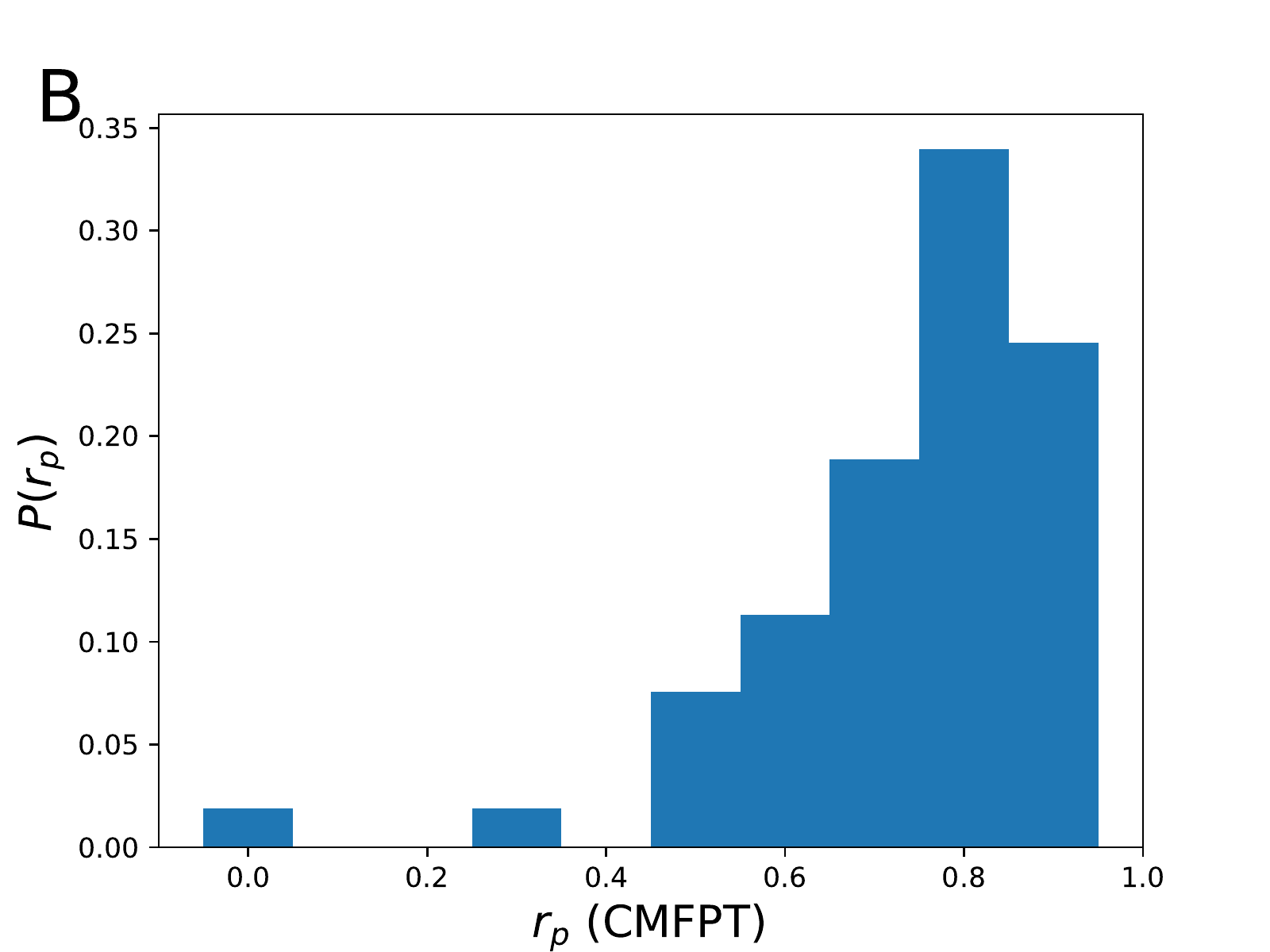}
    \end{center}
      \caption{\textbf{Correlation between $\widetilde{\tau}^k_{\rm sync}$ and the additional segregation indicators.} For each city in our study, we calculate the Pearson correlation coefficient $r_p$ between $\widetilde{\tau}^k_{\rm sync}$ and the additional segregation metrics over the $16$ income categories. The correlation coefficient for a city is thus obtained from a set of $16$ points, one per category. (A) Distribution of $r_p$ between Moran's I and $\widetilde{\tau}^k_{\rm sync}$ across cities. (B) Distribution of $r_p$ between the segregation calculated through normalized CMFPT $\mbox{med}(\widetilde{\tau})_k$ and $\widetilde{\tau}^k_{\rm sync}$ across cities.}
     \label{figrp}
\end{figure}

\begin{figure}[tb!]
    \begin{center}
    \includegraphics[width=0.45\textwidth]{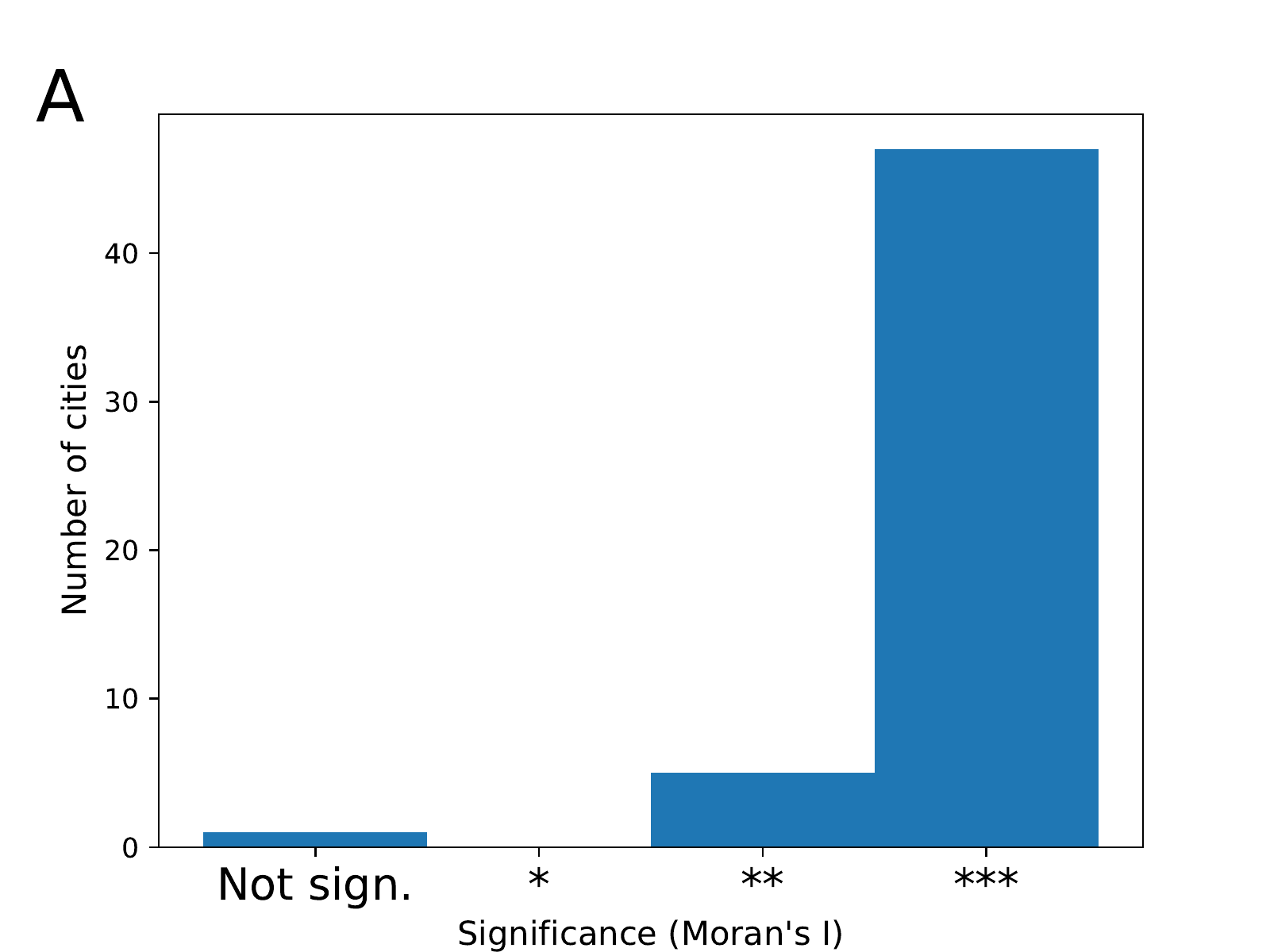}
    \includegraphics[width=0.45\textwidth]{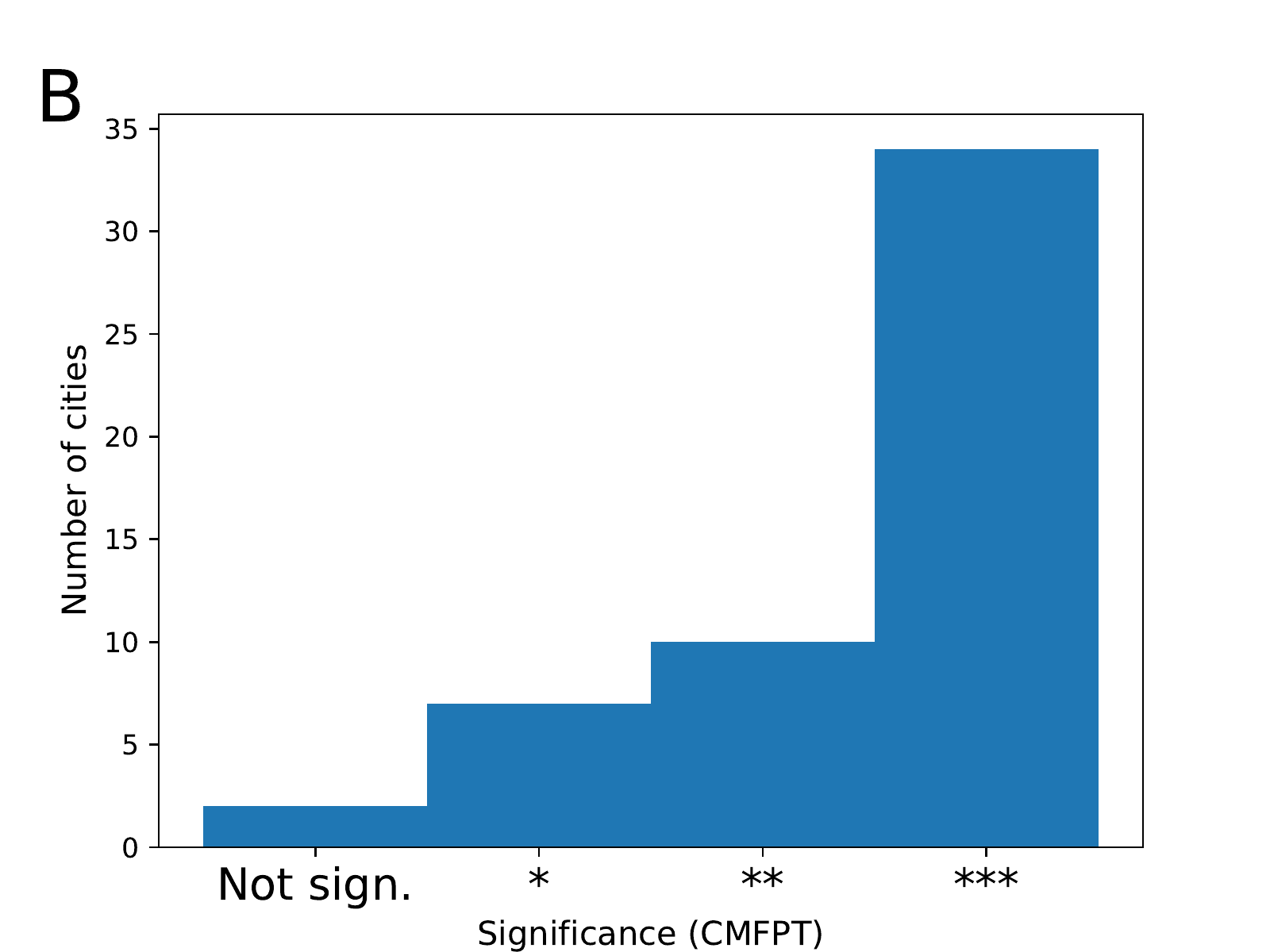}
    \end{center}
      \caption{\textbf{Significance of the Pearson correlation coefficients between $\widetilde{\tau}^k_{\rm sync}$ and the other segregation indicators.} For each of the additional indices, we display the significance of the correlations across cities. (A) Significance of correlations between Moran's I and $\widetilde{\tau}^k_{\rm sync}$. (B) Significance of correlations between the segregation calculated through normalized class mean first passage times $\mbox{med}(\widetilde{\tau})_i$ and $\widetilde{\tau}^k_{\rm sync}$ . The correlation coefficient and significance for each city is obtained by comparing the segregation values for the $16$ income categories. The significance values are depicted as * for $\mbox{p-value}<0.05$, ** for $\mbox{p-value}<0.01$, and *** for $\mbox{p-value}<0.001$.}
     \label{figsig}
\end{figure}

For each city included in our analysis, we measure the Pearson correlation coefficient $r_p$ between each of the additional segregation quantities and $\widetilde{\tau}^k_{\rm sync}$ for all the $16$ categories $k$. More specifically, for each city $r_p$ is calculated over a set of $16$ points. The distribution of $r_p$ across cities is shown in Fig.~\ref{figrp} for the Moran's I (A) and $\mbox{med}(\widetilde{\tau})_k$ (B), where a skewness towards high values is clearly observed. Most of the cities display correlations above $0.8$ with the Moran's I and $0.7$ with the CMFPT index. Additionally, we also show in Fig.~\ref{figsig} the significance of the correlations observed in each of the cities, which are also below $0.001$ in most of the cases.

\newpage

\begin{figure}[tb!]
    \begin{center}
    \includegraphics[width=0.90\textwidth]{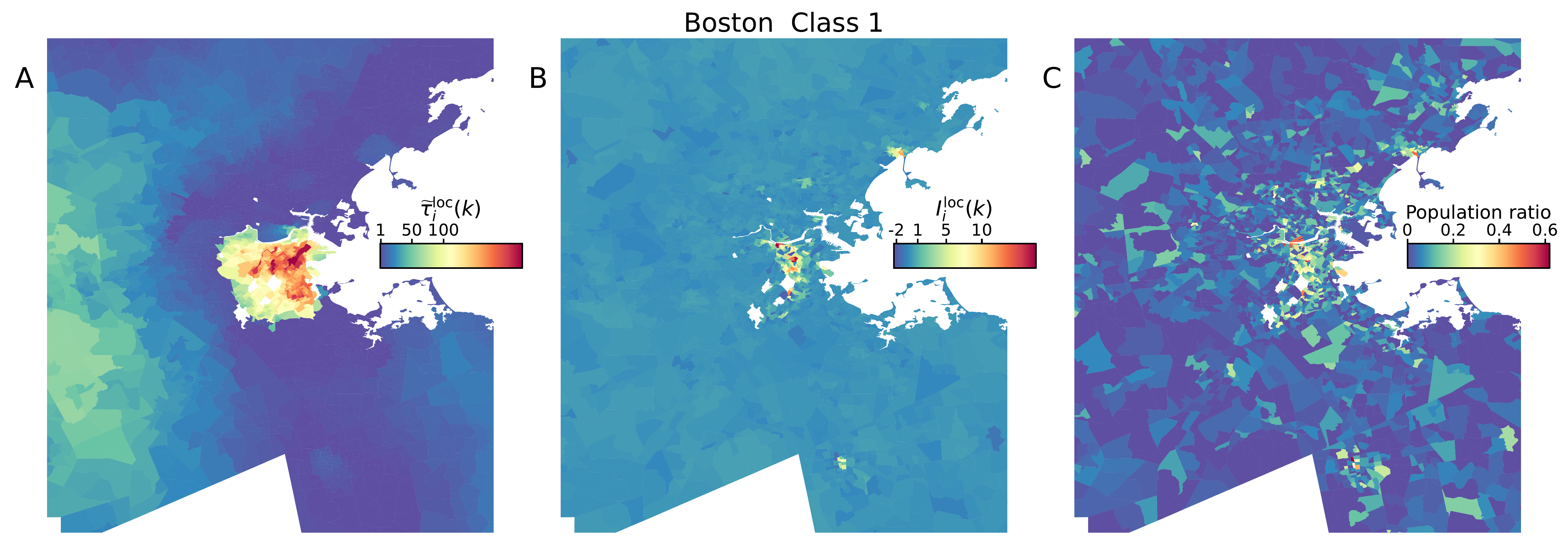} \\
    \includegraphics[width=0.90\textwidth]{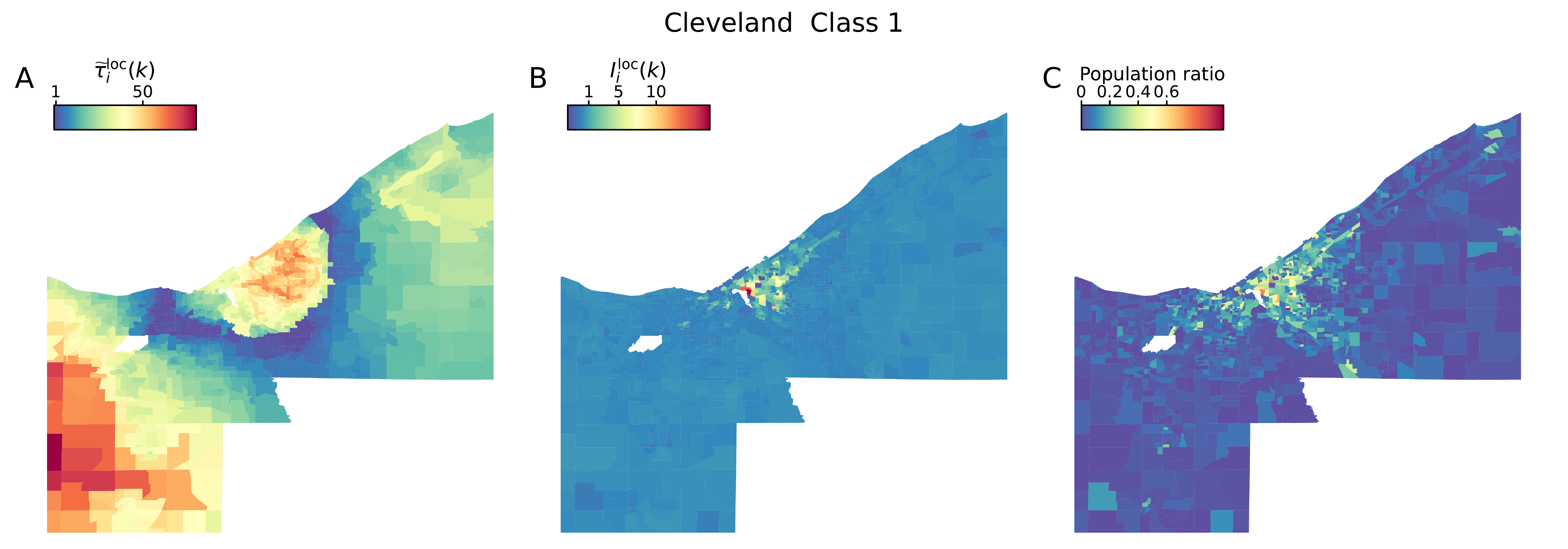} \\
    \includegraphics[width=0.90\textwidth]{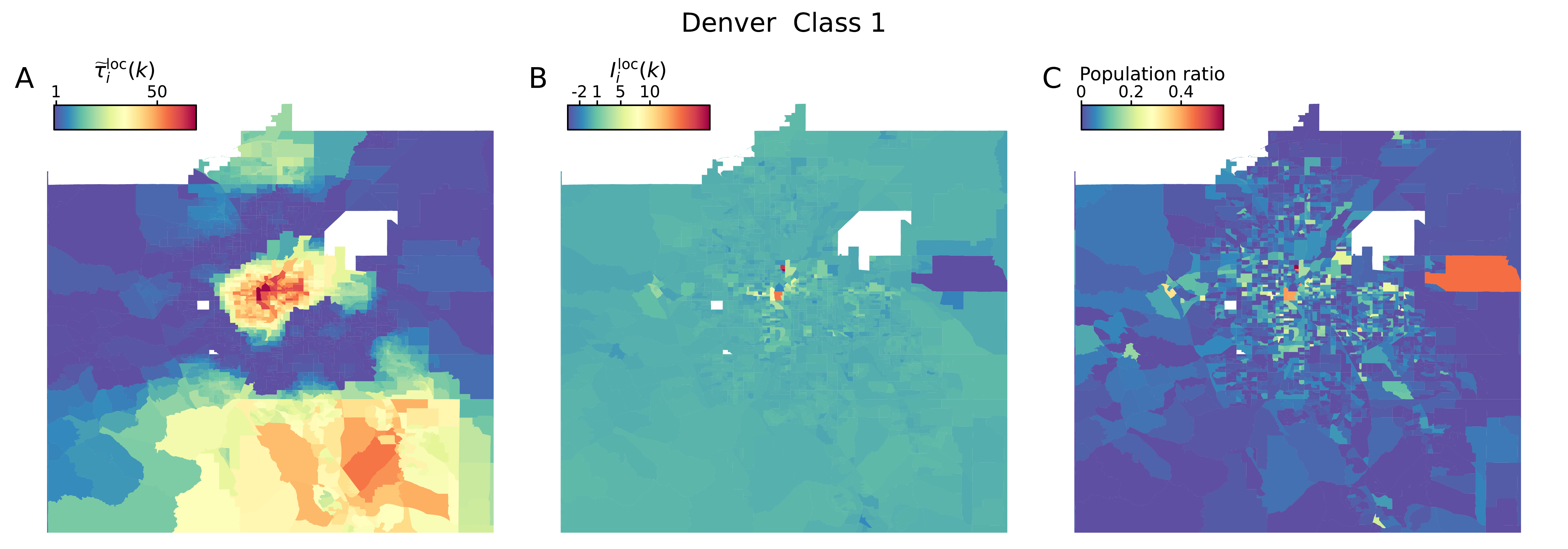} \\
    \includegraphics[width=0.90\textwidth]{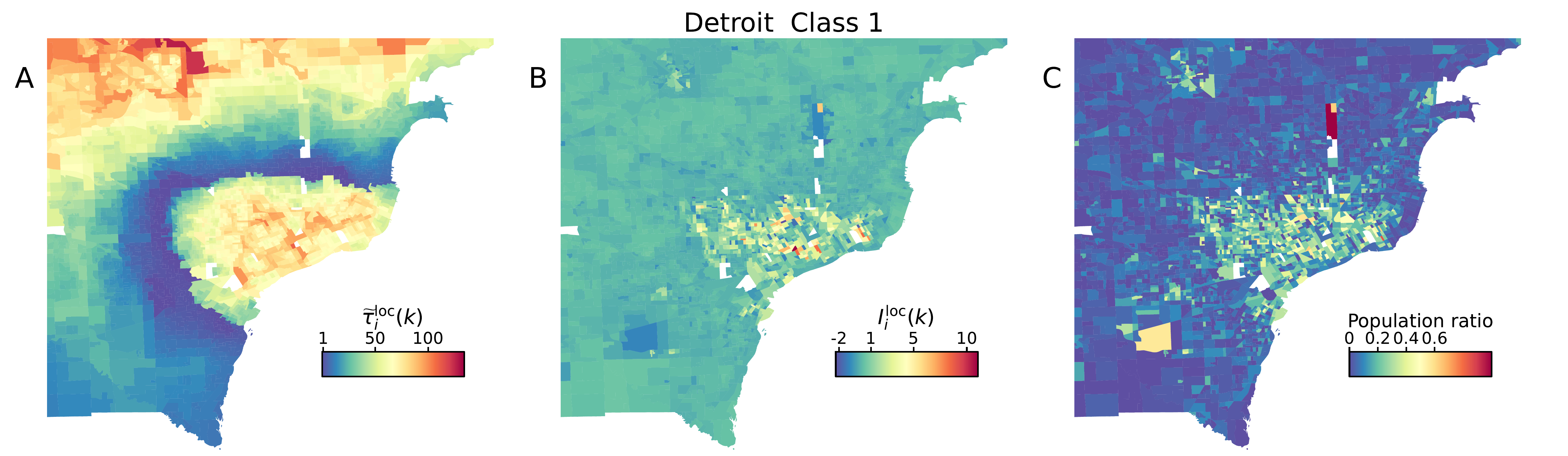}
    \end{center}
      \caption{\textbf{Comparison of local segregation indicators in Boston, Cleveland, Denver and Detroit.} (A) Normalized synchronization time, (B) Local Moran correlation, and (C) proportion of citizens for each census tract and income class~1 (most deprived).}
     \label{figcities}
\end{figure}

In the main text we discuss the potential of our methodology to assess the multiscale patterns of segregation in front of traditional first-neighbor approaches. In Fig.~\ref{figcities} we further investigate this fact by plotting for Boston, Cleveland, Denver and Detroit the local normalized synchronization times, the local Moran's $I^{\rm loc}_i(k)$, and the raw ratio of population of category $k$ in each of the census tracts.  

Although the segregation hotspots detected by our methodology and the local Moran's I seem similar, the patterns detected are significantly different. Whereas $I^{\rm loc}_i(k)$ captures strong differences between neighboors, $\widetilde{\tau}_i^{\rm loc}(k)$ highlights isolated regions even if the differences with their first-neighboors is low; most likely, this is because they are far apart from regions displaying ratios of population closer to the city average and require more time to reach the global synchronized state. In fact, the areas highlighted by synchronization dynamics have a larger scale and allow us to identify common mesoscale patterns of segregation across cities: a downtown that displays high values, a ring around it with low values, and finally the suburbs with high values again. By focusing on Detroit, we can see that not only the poorer downtown appears highlighted but also the suburbs due to their very low ratio of population of category $1$. Similar patterns can also be observed in Cleveland and Denver.

\begin{figure}[th!]
    \begin{center}
    \includegraphics[width=0.95\textwidth]{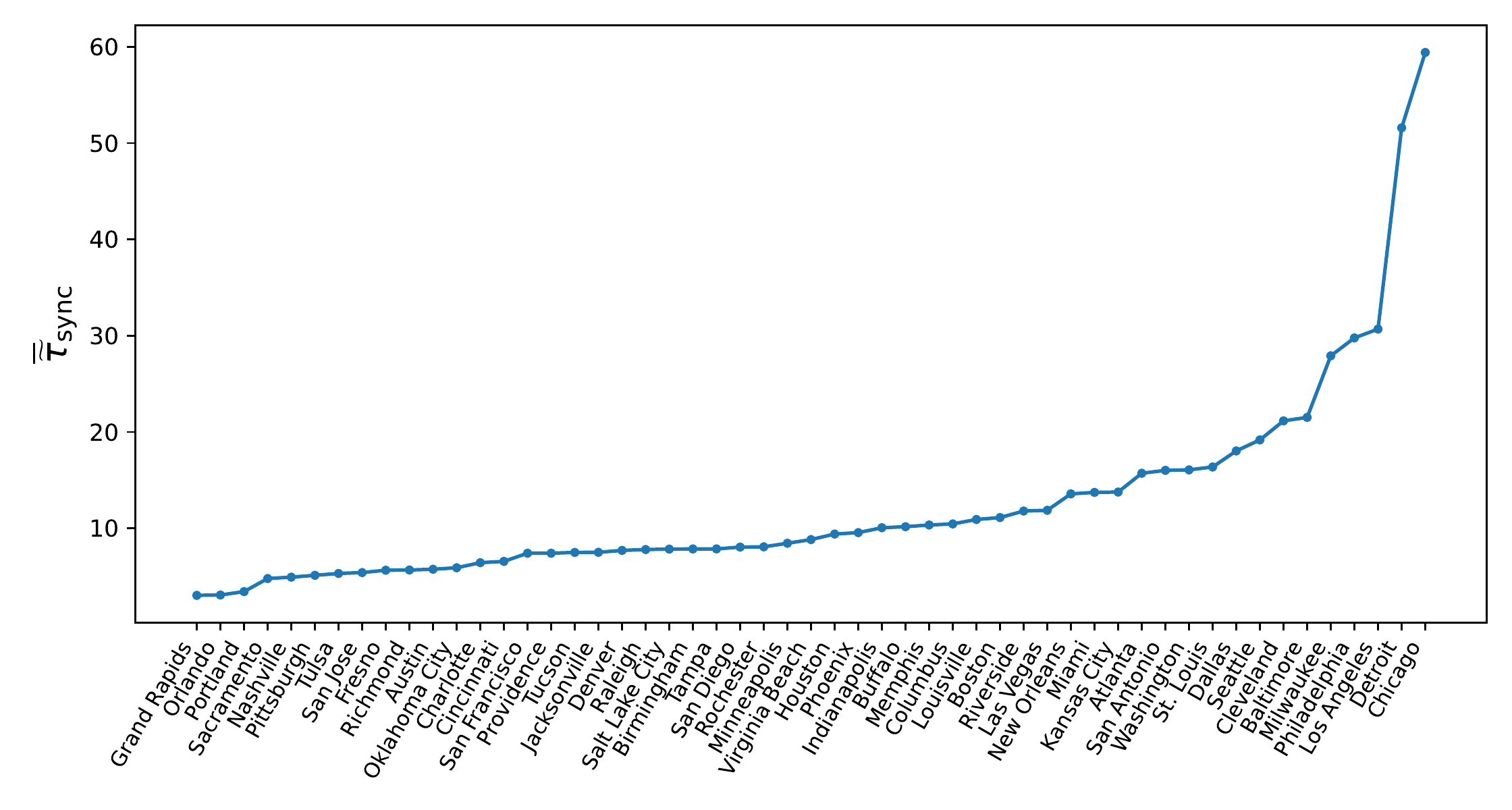}
    \end{center}
      \caption{\textbf{Segregation in US cities according to an index calculated through a weighted average.} Ranking of cities according to the weighted index of segregation $\bar{\widetilde{\tau}}_{\rm sync}$.}
     \label{figrank}
\end{figure}

\begin{figure}[th!]
    \begin{center}
    \includegraphics[width=0.55\textwidth]{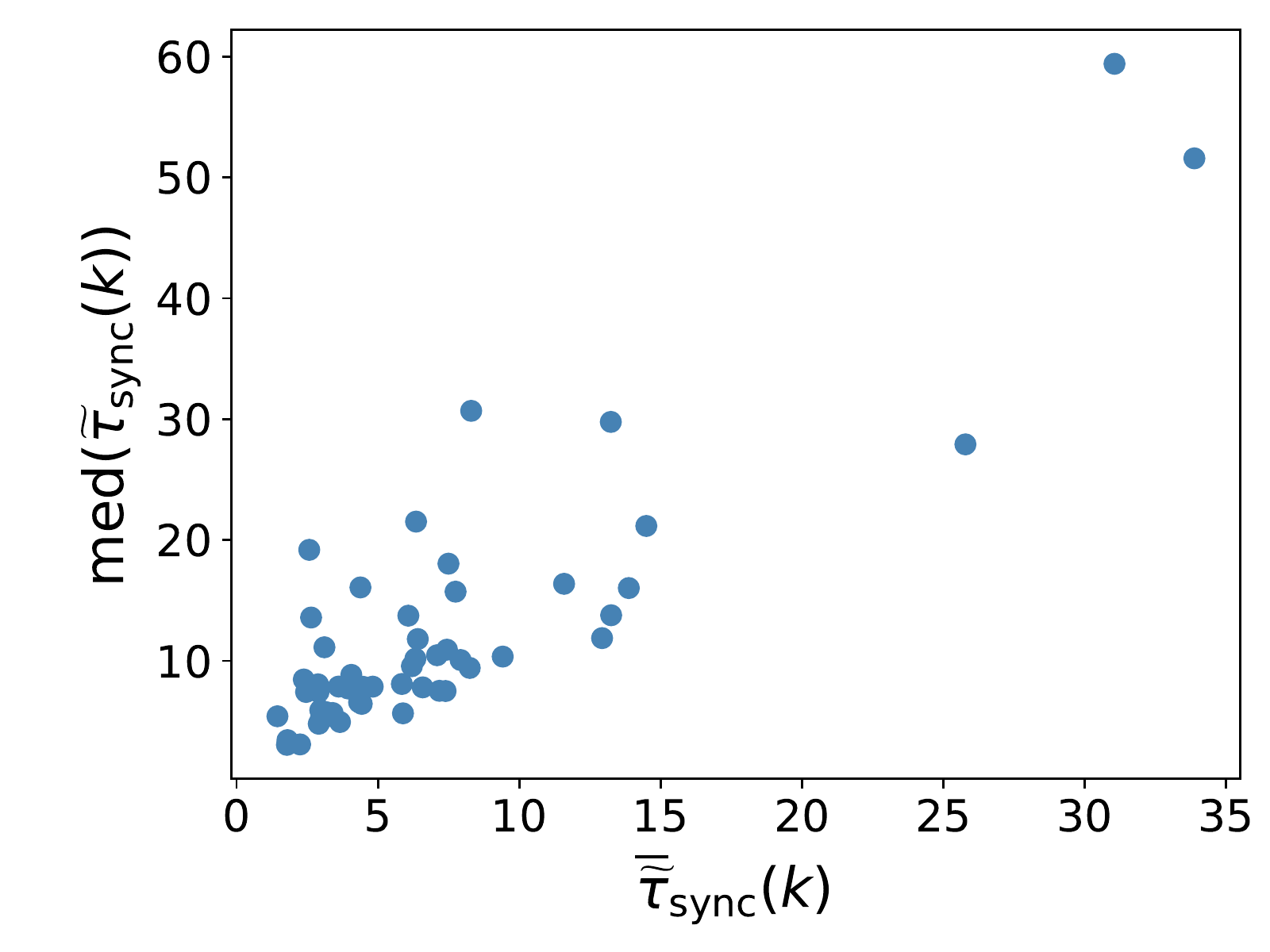}
    \end{center}
      \caption{\textbf{Comparison between segregation indicators obtained through synchronization dynamics.} Comparison between the weighted index of segregation $\bar{\widetilde{\tau}}_{\rm sync}$ and the index $\mbox{med}(\widetilde{\tau}_{\rm diff}(k))$ used in the main text.}
     \label{figcomp}
\end{figure}

The segregation index developed in the main text is calculated as the median of $\widetilde{\tau}^k_{\rm sync}$ which confers an equal weight to each of the income categories, disregarding the amount of population in each category. However, we can also construct a weighted index $\bar{\widetilde{\tau}}_{\rm sync}$ that can be built as
\begin{equation}
  \bar{\widetilde{\tau}}_{\rm sync}=\frac{\displaystyle\sum_k P_k \widetilde{\tau}_{\rm sync}(k)}{\displaystyle\sum_k P_k},
\end{equation}
where $P_k$ is the total number of citizens that belong to category $k$ in a given city. The ranking of cities according to the value of $\bar{\widetilde{\tau}}_{\rm sync}$ (Fig.~\ref{figrank}) displays only slight changes with, for example, Philadelphia and Los Angeles closer to the top of the ranking. We test the relation between both indices in Fig.~\ref{figcomp}, where a clear relationship between both quantities is revealed.

\clearpage
\section{Beyond economic segregation: Paris around the clock}

Besides only economic segregation, our methodology can be used to assess the spatial heterogeneity of any other quantity, and to exemplify it, we assess in this section the segregation of the population in Paris according to a wide set of socioeconomic indicators. The data compiles the fraction of population per district within a certain category at each hour of the day in French cities; in this work, we focus on Paris \cite{lecomte2018mobiliscope,julie2021,vallee2021}. The list of indicators and categories analyzed can be found in Table~\ref{TableS1}.

\begin{table}[ht!]
\centering
\begin{tabular}{llllll}
\hline \hline
Indicator & Categories \\
\hline
Activity type & At home & At work & Studying & Shopping & Leisure  \\
Age & 16-24 & 25-34 & 35-64 &   65 and more  &  \\
Educational level & Low  &  Middle-low &  Middle-high  &  High  & \\
Socioprofessional status  &  Inactive  &  Low  &   Middle-low  &   Middle-high &  High \\ 
Last travel mode  &  Pub. trans. &   Private motor  &  Soft mobility  &  &  \\
Occupational status  &  Active  &  Student  &  Unemployed  &  Retired  &  Inactive \\
Sex  &  Male  & Female  &  &  & \\ \hline
\end{tabular}
\caption{Socio-economic indicators and activity types analyzed for Paris.} \label{TableS1}
\end{table}

\begin{figure}[ht!]
    \begin{center}
    \includegraphics[width=0.90\textwidth]{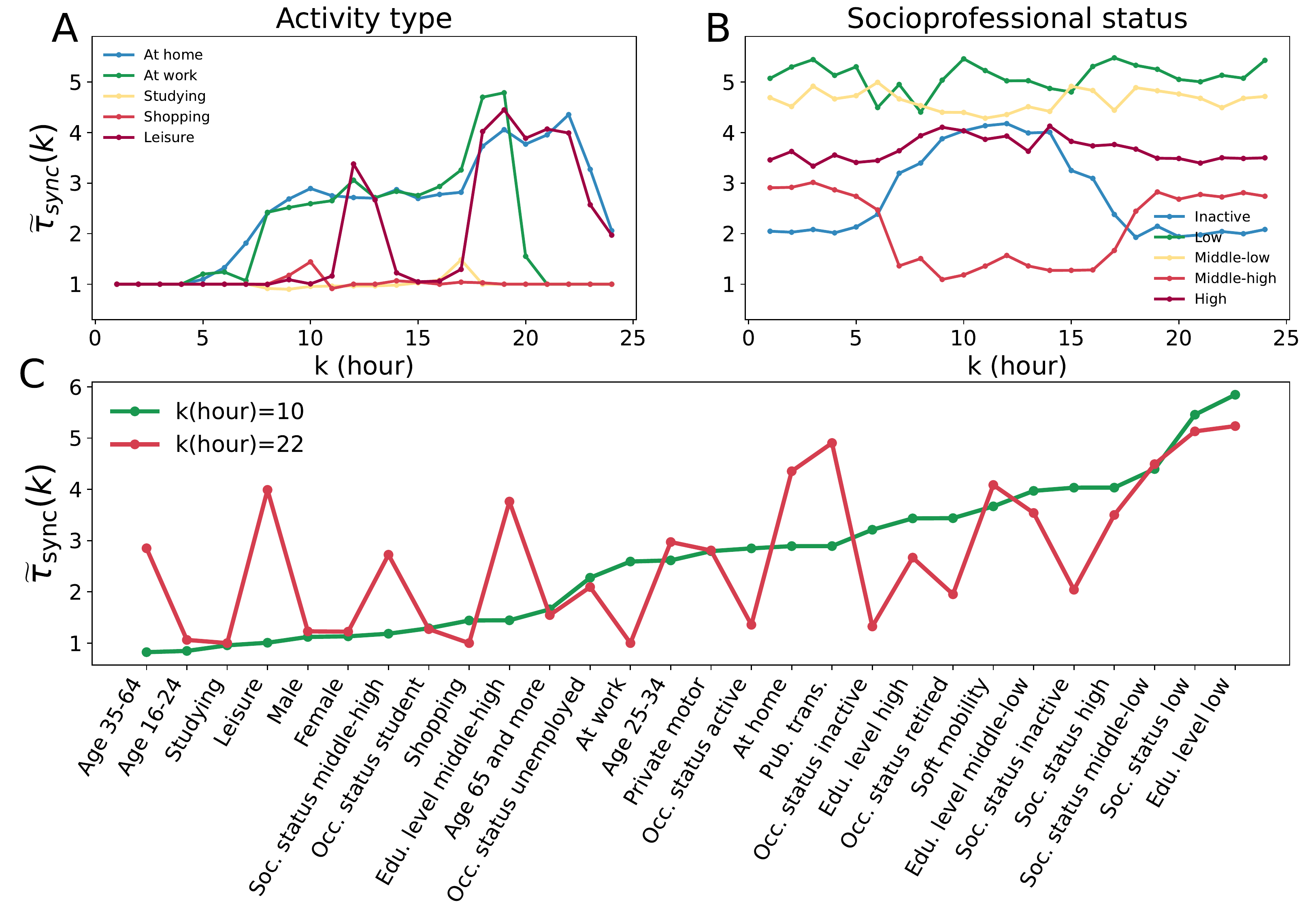}
    \end{center}
      \caption{\textbf{Synchronization around the clock in Paris.}  (A) Normalized synchronization time for the distribution of population performing each of the five types of activities. (B) Synchronization time for the distribution of population of each socio-professional status. (C) Change of synchronization times for all of the indicators at 10am (green) and 10pm (red).}
     \label{figS3}
\end{figure}

For each indicator or category, we have a certain distribution of population per spatial unit and hour of the day, thus we can compute how the quantity $\widetilde{\tau}_{\rm sync}(k)$
varies during the day, as we show in Fig.~\ref{figS3}(A,B) for the five activity types, and the five socio-professional status; the patterns of synchronization through time turn out to be very distinct.
For example, the level of synchronization remains basically constant throughout the day for low, middle and high socio-professional status, while it increases (decreases) between 8am and 8pm for inactive (high) socio-professional status. If we focus instead on the ranking of $\widetilde{\tau}_{\rm sync}(k)$ at 10am and 10pm, see Fig.~\ref{figS3}(C), the lower occupational and socio-professional status seem to be the most segregated indicators as they are on top of the ranking at both times of the day. Other categories that should be uniformly distributed across the city, such as sex, are very close to $1$, thus indicating no segregation.

\begin{figure}[ht!]
    \begin{center}
    \includegraphics[width=0.55\textwidth]{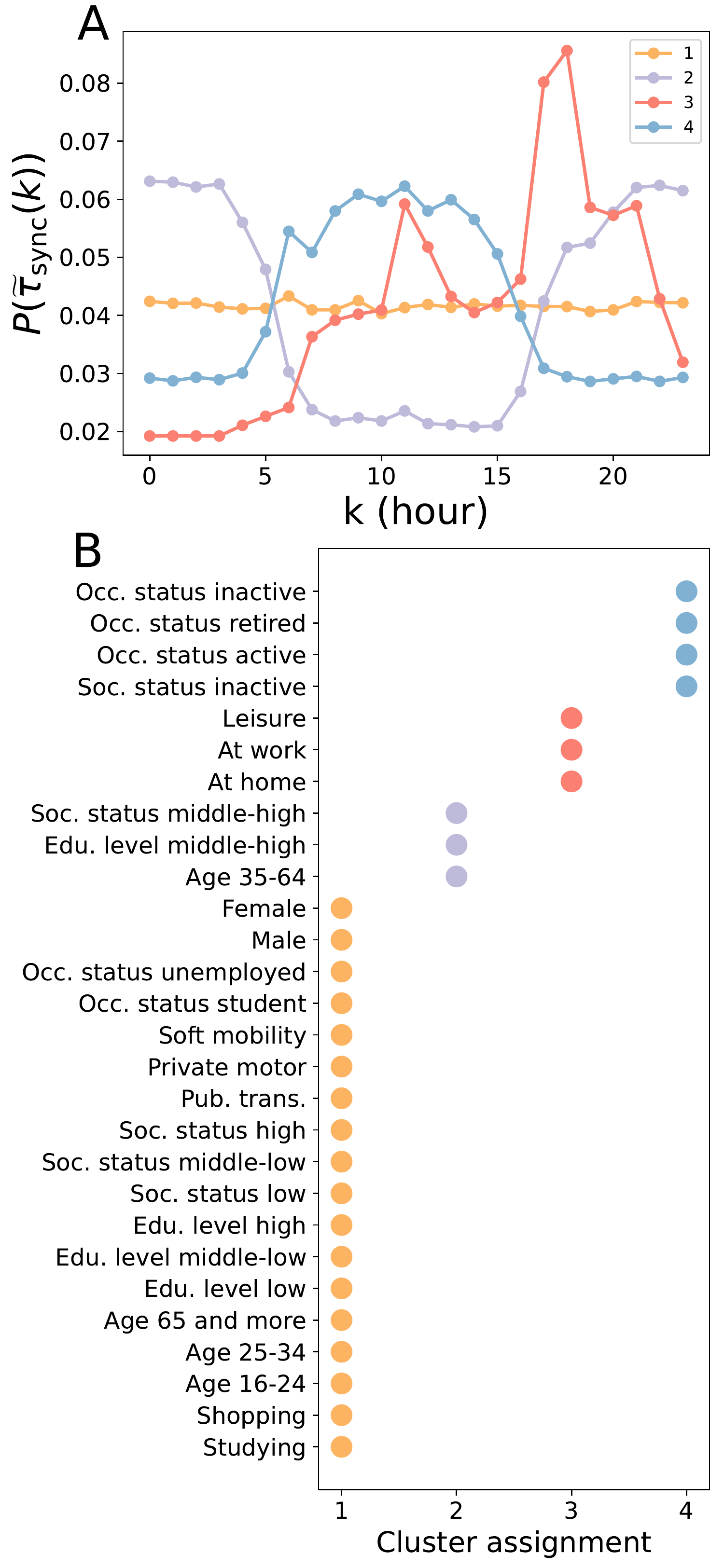}
    \end{center}
      \caption{\textbf{Clustering analysis of segregation around the clock in Paris.}  (A) Pattern of synchronization times $P$ for each of the four main groups detected with the K-Means algorithm. (B) Cluster assignment for each of the indicators analyzed.}
     \label{figS4}
\end{figure}

\begin{figure*}[ht!]
    \begin{center}
    \includegraphics[width=0.80\textwidth]{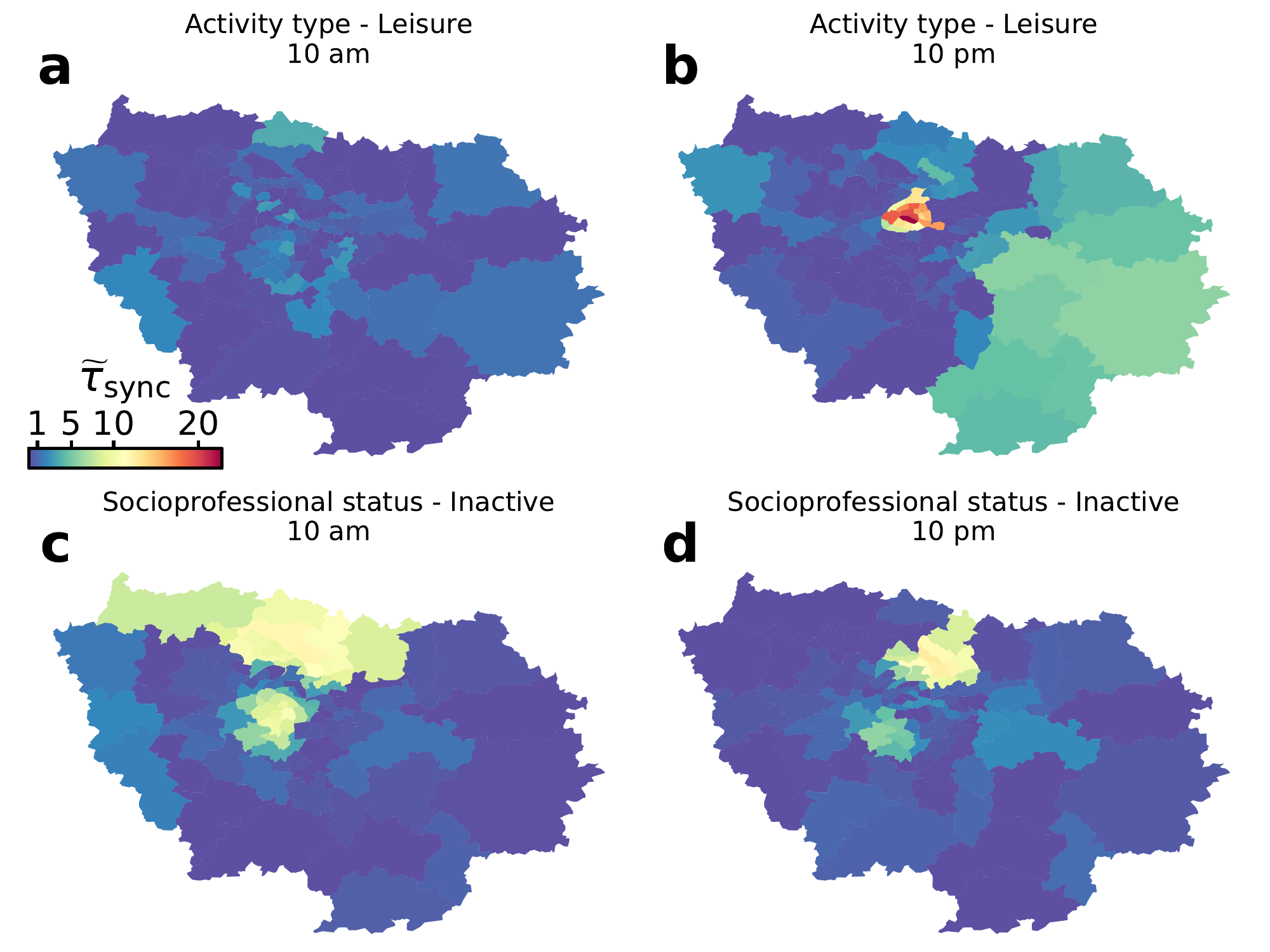}
    \end{center}
      \caption{\textbf{Local synchronization around the clock in Paris.} (A, B) Normalized synchronization time for each Paris district for the population performing leisure activities at 10am and 10pm. (C, D) Normalized synchronization time for each Paris district for the population with inactive socio-professional status at 10am and 10pm. For visualization purposes the color range is common to all four maps.}
     \label{figS5}
\end{figure*}

The hourly patterns of each metric allow for the grouping of indicators behaving similarly as we did for US cities. As before, we focus more on the time-series profile rather than the specific values taken by bon $\widetilde{\tau}^h_{\rm sync}(k)$, thus analyzing the normalized $P(\widetilde{\tau}^h_{\rm sync}(k))$ for each hour of the day $h$. The k-means clustering reveals four distinct clusters (see Fig.~\ref{figS4}) which correspond to: those increasing during workings, those decreasing, those remaining almost constant, and those with a more characteristic behavior with a peak during midday and at the end of the day, roughly around the lunch and dinner times.

Finally, we assess the local segregation of districts by measuring their local normalized synchronization time. In particular, we show an example in Fig.~\ref{figS5} for the population performing leisure activities and those with inactive socio-professional status.
In agreement with the temporal pattern shown in Fig.~\ref{figS3}, the segregation is much higher at 10pm compared to 10am, especially concentrated in the centre of the city; a not so surprising result given that most of the leisure activities are concentrated in that part of the city. In the case of the population with inactive socio-professional status, the hotspots seem to be concentrated in the northern part of the city, a region known for suffering a thriving inequality.

\bibliography{references}

\end{document}